\documentclass[journal,final]{IEEEtran}
\usepackage[utf8]{inputenc}
\usepackage{todonotes}
\usepackage{hyperref}
\usepackage{color,soul}
\usepackage{tabularx}
\usepackage[nomargin,inline,final]{fixme}
\fxusetheme{color}
\usepackage{multirow}
\usepackage{graphicx}
\usepackage{subcaption}
\usepackage{enumitem}
\usepackage{rotating}
\usepackage{float}
\usepackage{adjustbox}
\usepackage{booktabs}
\usepackage{textcomp}

\FXRegisterAuthor{ra}{ara}{ra}
\FXRegisterAuthor{pf}{apf}{pf}
\FXRegisterAuthor{fs}{afs}{fs}
\FXRegisterAuthor{nb}{anb}{nb}

\FXRegisterAuthor{rv}{rvc}{rv}

\title{Visual Distortions in 360-degree Videos}

\author{
  Roberto~G.~de~A.~Azevedo, 
  Neil~Birkbeck, 
  Francesca~De~Simone, 
  Ivan~Janatra, 
  Balu~Adsumilli, 
  Pascal~Frossard 
\thanks{R. Azevedo and P. Frossard are with
        Signal Processing Lab.~(LTS4),
        École Polytechnique Fédérale de Lausanne~(EPFL),
        Lausanne, Switzerland.}%
\thanks{F. de Simone is with
        Centrum Wiskunde \& Informatica~(CWI),
        Amsterdam, The Netherlands.}%
\thanks{N. Birkbeck, I. Janatra, and B. Adsumilli are with
        Youtube,
        Mountain View, California, USA.}}

\date{\today}

\begin{document}

\maketitle

\begin{abstract}
Omnidirectional~(or 360-degree) images and videos are emergent signals in many
areas such as robotics and virtual/augmented reality.
In particular, for virtual reality, they allow an immersive experience in
which the user is provided with a 360-degree field of view and can navigate
throughout a scene, e.g., through the use of Head Mounted Displays.
Since it represents the full 360-degree field of view from one point of the
scene, omnidirectional content is naturally represented as spherical visual
signals.
Current approaches for capturing, processing, delivering, and displaying
360-degree content, however, present many open technical challenges and
introduce several types of distortions in these visual signals.
Some of the distortions are specific to the nature of 360-degree images, and
often different from those encountered in the classical image communication
framework. 
This paper provides a first comprehensive review of the most common visual
distortions that alter 360-degree signals undergoing state of the art
processing in common applications.
While their impact on viewers' visual perception and on the immersive
experience at large is still unknown ---thus, it stays an open research
topic--- this review serves the purpose of identifying the main causes of
visual distortions in the end-to-end 360-degree content distribution pipeline.
It is essential as a basis for benchmarking different processing techniques,
allowing the effective design of new algorithms and applications.
It is also necessary to the deployment of proper psychovisual studies to
characterise the human perception of these new images in interactive and
immersive applications.

\end{abstract}

\begin{IEEEkeywords}
Omnidirectional video, 360-degree video, visual distortions, artifacts,
  compression.
\end{IEEEkeywords}

\section{Introduction}
\label{sec:introduction}

\IEEEPARstart{F}{rom} Virtual Reality~(VR) to robotics, innovative
applications exploiting omnidirectional images and videos are expected to
become widespread in the near future.
Fully omnidirectional cameras, able to capture a 360-degree real-world scene,
have recently started to appear as commercial products 
and professional tools. 
User-generated and professional 360-degree content is already being
distributed using popular content sharing platforms, such as YouTube
and Facebook.

While the popularity of 360-degree content and applications is rapidly
increasing, many technical challenges at different steps of the
omnidirectional signal acquisition, processing, and distribution chain remain
open.
The current approaches to process and distribute omnidirectional visual
signals rely on algorithms and technologies designed for classical image and
video signals captured by perspective cameras.
However,
the omnidirectional imaging pipeline has some particularities that induce
specific distortions if not handled properly.
First, the geometry of the content capture system is spherical, rather than
planar~\cite{Micusik2004}.
To reuse existing file formats and algorithms designed for perspective
signals, the spherical signal is warped into a planar
representation~\cite{Chen2018}.
The resulting planar signal, however, is not a classical natural visual
signal~\cite{DeSimone2016}.
Second, the 360-degree content rendering, eventually via a Head Mounted
Display~(HMD), is characterised by an interactive and immersive dimension that
represents a significant novelty~\cite{petry2015}.
Therefore, existing algorithms need to be adapted and optimized to process
these signals efficiently and satisfy the new requirements.

In addition, 360-degree content distribution is expected to push the current
storage and network capacities to their limits.
Most HMDs currently available on the market provide up to full High
Definition~(HD) display resolution.
Since these devices usually provide a 110-degree of field-of-view, thus 4K
resolution is being widely accepted as a minimum functional resolution for the
full 360-degree planar signal.
Nevertheless, HMDs with 4K or 8K display resolution are already appearing on
the market.
Thus, 360-degree planar signals with resolution of 12K or higher will have to
be efficiently stored and transmitted soon~\cite{hosseini2017view-aware}.
Consequently, new coding and transmission techniques, able to cope with
increasingly high data rates and to satisfy user's expectations regarding
visual quality, will be continuously needed.

To improve existing omnidirectional processing pipelines and design new
perceptually-optimised omnidirectional visual communications, it becomes
critical to design tools to detect the \emph{visual distortions}~(or
\emph{artifacts}) introduced by each processing step, and, ultimately,
quantify their impact on the perceived quality of the signal presented to the
user and on the immersive experience~\cite{desimone2017omnidirectional}.

Visual distortions occurring in images and videos captured by perspective
cameras and undergoing compression and transmission have been largely
characterized and analyzed in the literature, both for standard 2D
~\cite{yuen1998survey,wu2005coding,unterweger2013compression,zeng2014characterizing}
and stereoscopic 3D
signals~\cite{meesters_survey_2004,boev2008classification,boev2009stereoscopic,hanhart20133d}.
Nevertheless, \emph{new types of distortions} can occur in 360-degree visual
signals dataflows.
%
These distortions
have not been characterized in the literature yet, to the best of the authors'
knowledge. 
In~\cite{knorr2017modular}, a classification of the distortions caused by
360-degree content capture is presented, but the analysis does not include
other processing steps such as coding, transmission, rendering, as well as the
impact of the display technology.
Some works reporting upon subjective studies in which users are asked to
assess the overall quality of a set of processed 360-degree images or videos
have recently appeared in the literature~\cite{%
  ebrahimi2017measuring,
  schatz2017towards,
  duan2018perceptual,
  lim2018vr,
  xu2017visual,
  zhang2018subjective,
  zou2018perceptual}.
These works focus on the methodology used to collect user feedback and
provide valuable guidelines to deploy classical subjective quality assessment
experiments, but none of them analyzes the perceptual impact of 360-degree
specific distortions.



\begin{figure*}[!htb]
  \centering
  \includegraphics[width=.8\textwidth]{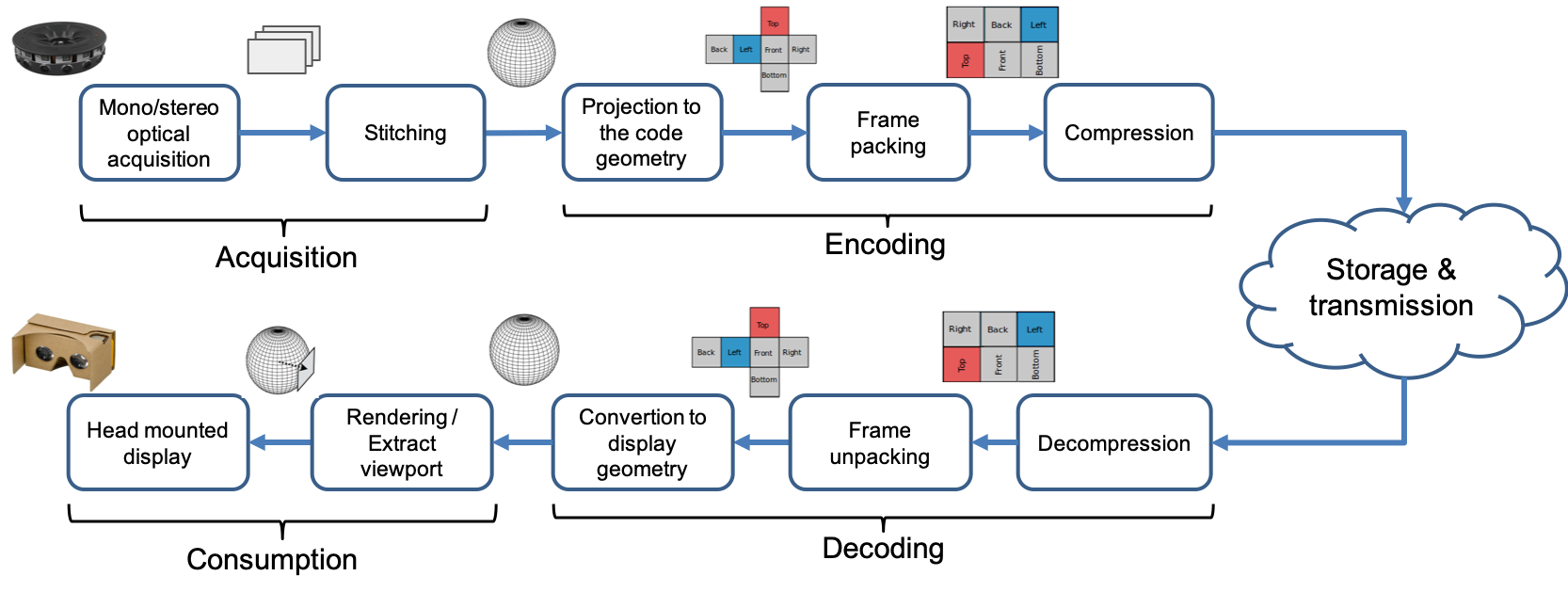}
  \caption{End-to-end 360-degree video processing pipeline.} 
  \label{fig:360video_pipeline}
\end{figure*}

This paper reviews and characterizes the most common visual distortions found
in 360-degree signals undergoing state of the art end-to-end processing,
including acquisition, lossy compression, transmission, and visualization by
the end user.
The goal is the isolation of individual distortions with the aim of obtaining
a description of their visual manifestations, causes, and relationships.
Visual examples are presented when possible\footnote{
  The visual examples presented in this paper are mainly to illustrate the
  visual effect of the distortions.
  A true visual appreciation of the artifacts can only be provided by viewing
  the affected sequences on an HMD.}.
This timely review serves
as a tool 
for the benchmark of different processing algorithms and display devices, in
terms of perceptual quality and will help in the deployment of psychovisual
studies to characterize human perception of these new signals in new
consumption scenarios.
Moreover, being aware of the different visual artifacts and their causes is
necessary for the development of more effective algorithms, that are able to
properly cope with the specific nature of 360-degree images. 
Also, a brief overview of the existing tools used in state of the art to
assess the quality of omnidirectional signals is presented, and perspective on
future research directions are discussed.

The remainder of the paper is organized as follows.
Section~\ref{sec:360pipeline} presents the typical 360-degree video processing
pipeline, used by most of the state-of-the-art approaches, and reviews some of
the processing techniques currently in use in each step.
Sections~\ref{sec:acquisition}--\ref{sec:display} detail, for each step of the
360-degree\ pipeline, from acquisition to visualization, the most common
artifacts and discuss their causes.
Section~\ref{sec:discussion} discusses the current approaches, the open
issues on the visual quality assessment of 360-degree videos, and how to use
this comprehensive review to improve them.
Finally, Section~\ref{sec:conclusion} brings our conclusions and points out
future work.

\section{360-degree video processing pipeline}
\label{sec:360pipeline}

Fig.~\ref{fig:360video_pipeline} depicts the end-to-end 360-degree signal
processing pipeline that we consider in this paper, from acquisition to
consumption by the end user via an HMD.
Each step is briefly described hereafter.

\subsection{Acquisition}

Different optical systems have been proposed in the past to \emph{capture}
wide field of view
signals~\cite{yagi1999omnidirectional,gurrieri2013acquisition}.
Nowadays, most of the commercial omnidirectional cameras with a full
360-degree field of view~(e.g., the Ricoh Theta, the Gear360, and the Orah
cameras) are multi-sensor systems, in which each sensor is a dioptric camera
(sometimes with fish-eye lenses).
These systems can be modeled as central cameras that project a point in the 3D
space to a point on a spherical imaging surface, i.e., the viewing
sphere~\cite{Micusik2004}.
In practice, the omnidirectional output signal 
is the result of a \emph{mosaicking}~(i.e., \emph{stitching}) algorithm,
specific to the acquisition systems, which merges the
overlapping field of view signals acquired by all dioptric sensors to produce
a wide-view \emph{panorama} image~\cite{szeliski2007image}.

In automatic \emph{image stitching} processes, the overlapping regions
between the cameras are \emph{aligned} using different planar models~(e.g.,
affine, perspective, or cubic transformation models); then, the views are
\emph{blended} and \emph{warped} to the omnidirectional 3D surface, commonly a
sphere surface~\cite{knorr2017modular}.
For \emph{video stitching}, additional \emph{video synchronization}~(if the
individual sensors are not finely synchronized) and \emph{video
stabilization}~(for moving cameras) may be
necessary~\cite{xu2012panoramic,jiang2015video}.

The output signal of the stitching process is usually stored, using standard
file formats, as a rectangular array of samples (\emph{planar
representation}), resulting from the projection of the sphere to a plane
(\emph{map projection} or \emph{spherical
parametrization})~\cite{pearson1990map}.
The planar representation allows re-using existing image and video content
distribution chains, including encoders, packagers, and transmission
protocols.
Additionally, it is practical for rendering since hardware graphics systems
need a simple arrangement of samples to access spherical images as a texture
map.
Most of the consumer-level omnidirectional cameras stitch to a planar
representation referred to as \emph{equirectangular
panorama}~(Fig.~\ref{fig:map_projections_erp}).
(Some professional-level cameras also allow to access the individual camera
input, so that it is possible to use off-line software or manually fix some
of the stitching problems, which are discussed later).
The panorama uses an equirectangular projection~(ERP) that maps a sphere to a
plane by sampling the spherical signal on an equi-angular grid and using the
longitude and latitude of each sample on the sphere as coordinates of the
sample projected on the plane~\cite{pearson1990map}.

A few stereoscopic omnidirectional camera systems, able to capture the stereo
views in all directions~\cite{peleg2001omnistereo}, have also been recently
built as prototype~\cite{schroers2018omnistereoscopic} and professional
capture systems ---e.g., Facebook Surround 360, Jump~\cite{anderson2016jump},
Obsidian~\cite{tan2018360-degree}.
They commonly output an Omni-Directional Stereo~(ODS) representation that
contains two modified ERP signals~\cite{ishiguro1992ods}, corresponding to the
left and right views for the human eyes.
Capturing stereoscopic omnidirectional dynamic scenes, however, is very
challenging, since there is the inherent problem of self-occlusion among the
cameras.
A broad discussion on the different possibilities for the acquisition of both
static and dynamic omnidirectional stereoscopic content is provided
in~\cite{gurrieri2013acquisition}.

%



\subsection{Encoding}

The goal of the encoding step is to reduce, in a lossless or lossy way, the
redundancy in the signal, and thus the space needed to store and transmit it.
Most of the omnidirectional video systems re-use the same encoding tools as
classical video solutions, such as H.264, H.265, VP9, or AV1~\nbnote{ref?}.
The main challenge with omnidirectional coding resides in mapping the content
into rectangular frames that are typical inputs for these video encoders.

A straightforward solution to encode 360-degree visual signals is to directly
use the ERP~(or ODS) signal output by an omnidirectional camera as input for
any state-of-the-art encoder.
Nevertheless, the equirectangular representation is not the most efficient
representation for encoding.
First, the regular sample distribution in the planar domain corresponds to a
non-uniform sampling density on the sphere, with higher density towards the
polar areas.
Such a sample distribution is wasteful
because, as have been demonstrated by subjective tests and head motion
capturing study~\cite{wu2017dataset,corbillon2017360-degree}, the content at
the poles is usually not the most semantically interesting part of the scene
being captured.
Second, the ERP signal presents strong warping distortions towards the top and
bottom image boundaries, which correspond to the polar areas in the spherical
domain.
Besides these geometric properties, the omnidirectional image signal has
statistical characteristics which are not those of typical natural visual
signals generated by perspective cameras, for which the encoding tools have
been tuned for.
By using the ERP representation in classical video encoders, the compression
is therefore suboptimal.

\begin{figure*}[!htb]
  \centering
  \begin{subfigure}[b]{.75\textwidth}
    \includegraphics[width=\textwidth]{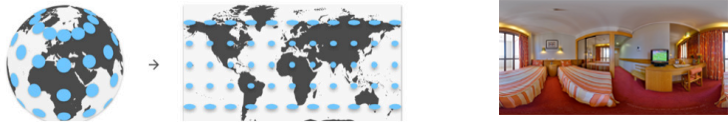}
    \caption{Equirectangular projection~(ERP). ERP maps meridians to equally
      spaced vertical straight lines, and parallels to equally spaced
      horizontal lines.  In this projection, poles regions are stretched
      compared to the equator region.}
    \label{fig:map_projections_erp}
  \end{subfigure}
  \par\bigskip 
  \begin{subfigure}[b]{.7\textwidth}
    \includegraphics[width=\textwidth]{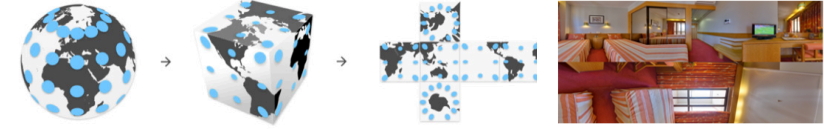}
    \caption{Cubemap projection (CMP). CMP is performed by projecting the
      sphere signal to a circumscribed cube.  It results in six faces that
      need to be rearranged to form a rectangular frame. \ranote{Sampling?}}
    \label{fig:map_projections_cmp}
  \end{subfigure}
  \caption{Examples of map projections.}
  \label{fig:map_projections}
\end{figure*}

Alternative planar representations that address both problems, by implying a
more uniform sampling density in the spherical domain and being characterized
by less strong warping distortions, have been proposed in the literature, such
as \emph{cube map}~(CMP), \emph{octahedron}, and tile-based
projections~\cite{chen2018recent}.
Among those, CMP is the most common one.
CMP is composed by the projection of the sphere in a circumscribed cube,
resulting in six square cube faces~(see Fig.~\ref{fig:map_projections_cmp}).
Some studies have shown that using CMP can save up to 25\% of the
bitrate when compared to a similar user perceived quality in the ERP
format~\cite{facebook2016cubemap_blog}.
Also, CMP is well-known in the computer graphics and gaming communities, and
thus it is well-supported by graphics frameworks such as
OpenGL~\cite{GL44spec}.

As exemplified by the CMP format, some of the current projection methods
result in different sets of faces, which then need to be packed together into
one planar image~(\emph{frame packing} step of
Fig.~\ref{fig:360video_pipeline}).
For instance, a common packing method for CMP is the cubemap 3x2 arrangement,
shown in the sample content of Fig.~\ref{fig:map_projections} (right image).
Different frame packing methods may result in different discontinuities
between the faces.
A primary goal of the frame packing is to minimize the number of
discontinuities in the planar representation.

Once the arrangement of faces has been completed, rectangular frames are
constructed, possibly with additional padding, and eventually fed into
classical video compression engines. 

In the case of stereoscopic omnidirectional content, the individual
omnidirectional images for each eye are usually packed together in a
frame-compatible stereo interleaving approach~\cite{dufaux20133d}, e.g.,
through a top/bottom or side-by-side frame representation.~\ranote{Maybe we
could add an ODS representation in Fig.~\ref{fig:map_projections}.}
In theory, other approaches that have been explored for standard stereoscopic
3D video ---such as simulcast, asymmetric coding, multiview coding,
etc.~\cite{dufaux20133d}--- can also be adapted to the omnidirectional
stereoscopic case.
However, since such approaches are still underexplored for 360-degree content,
they are not considered in the rest of the paper.  

\subsection{Transmission}
\label{subsec:transmission}

In principle, since the encoding process results in traditionally compressed
2D or stereoscopic 3D frames, 360-degree delivery can use the same video
streaming algorithms as classical image communication systems.
Nevertheless, 360-degree content implies new technical challenges on content
distribution due to the high data rate of omnidirectional signals and the low
latency requirements of immersive communication.
In addition, unlike conventional video, the user does not look at the entire
scene at once and can navigate around the content.

Nowadays, to reuse existing delivery architectures for video on demand and
live streaming services, content delivery solutions relying on Dynamic
Adaptive Streaming over HTTP~(DASH)~\cite{stockhammer2011dynamic} are the most
prominent ones to 360-degree video~%
\cite{hosseini2016adaptive,
      elganainy2016streaming,
      sreedhar2016viewport,
      ozcinar2017viewport,
      corbillon2017viewport,
      graf2017towards,
      xie2017360ProbDASH,
      liu2017joint,
      ghosh2017rate,
      zhou2017measurement,
      timmerer2017adaptive,
      concolato2017adaptive,
      zhou2018effectiveness}.
In these approaches, the server stores an adaptation set, i.e., a set of
multiple versions~(representations) of the same content, encoded at different
bit-rates and resolutions.
Each representation is temporally divided into consecutive segments of fixed
duration, commonly ranging from 1s to 5s.


Since only a portion of the whole spherical video~(the \emph{viewport}) is
displayed by the HMD at any given time instant, a number of
\emph{viewport-aware} streaming schemes have been recently devised to exploit
this fact~(in contrast to the \emph{viewport-agnostic ones}, which handle the
omnidirectional video as a conventional 2D video).
In such an approach, a client is provided with different video
representations, each one favoring a different viewport.
The client should select and download some of the representations by taking
into account not only the prediction of the available bandwidth but also a
prediction of the user's navigation pattern.
Nowadays, two main variations of viewport-aware dynamic adaptive streaming are
being explored: \emph{viewport-dependent projection} and \emph{tile-based
streaming}.

In the \emph{viewport-dependent projection}
approach~\cite{%
  sreedhar2016viewport,
  corbillon2017viewport,
  zhou2017measurement,
  zhou2018effectiveness}
a certain area may be favored in the planar representation using different
projections.
It is possible to have different viewport-dependent quality representations,
each one favoring a specific viewport of the content.
Thus, the client can choose the optimal viewing quality approach by selecting
the projection representation which provides the best representation of the
current user's viewport.

In the \emph{tile-based} approach~\cite{%
  hosseini2016adaptive,
  ozcinar2017viewport,
  graf2017towards,
  xie2017360ProbDASH,
  liu2017joint,
  ghosh2017rate,
  timmerer2017adaptive,
  concolato2017adaptive,
  ozcinar2017estimation}
the planar omnidirectional video is decomposed into independently decodable
rectangular parts, i.e., tiles, so that each tile is encoded at different
quality levels.
The client can then choose to download only the tiles contained in the current
user's viewport with high quality while downloading the non-visible ones with
lower quality, or even ignoring them.

\subsection{Consumption}

At the client side, the inverse steps ---\emph{decoding}, \emph{unpacking},
\emph{conversion to display geometry} and \emph{viewport extraction}~(or
\emph{rendering})--- need to be performed, so that the user can visualize and
interact with the 360-degree video content.
When the content is rendered to be visualized, the inverse mapping from the
plane to the sphere is performed.
The viewer is centered on the sphere and is able to navigate the content by
changing their viewing direction: a portion of the sphere surface is projected
to the \emph{viewport}, depending on the user's viewing direction.
This rendering is typically implemented on a Head-Mounted Display (HMD), or on
more classical devices such as a computer or smartphone.
With HMDs, the users can easily navigate the scene by turning their head
freely.
In desktops and smartphones, the users can consume the 360-degree content
through a ``magic window'', in which they can interact with the content using
a mouse~(or another device) in a desktop, or by moving the position of the
smartphone in the physical space.
Given the more immersive features and challenging of HMD-based approaches
(i.e., it can introduce new distortions still not fully understood), with
regards to consumption, this paper focus mainly on the HMD-based approaches.

\section{Artifacts caused by Acquisition}
\label{sec:acquisition}


As previously mentioned, capturing 360-degree content is usually composed of
two main steps: acquiring the visual content through a multicamera optical
system and then stitching the multiple images into one global signal,
generally in the form of a spherical image.
Each of these steps may add visual distortions, which are discussed in what
follows.


\subsection{Sensor limitations}

Today, the most common omnidirectional recording systems are composed of
multiple cameras, which altogether capture omnidirectional views of real-world
scenes.
Each of the cameras of a 360-degree multicamera rig is subject to common
\emph{optical distortion} ---e.g., \emph{barrel}, \emph{pincushion
distortions}, and \emph{chromatic aberrations}--- \emph{moiré effect},
\emph{noise}, and \emph{motion blur}~\cite{boev2008classification}.
In particular, wide-angle fisheye cameras, commonly used in the multicamera
rigs, are prone to \emph{chromatic aberrations}, more than regular perspective
cameras.
In addition, wide aperture angles on fish-eye cameras are only possible with
large amounts of \emph{barrel distortion}.

Additional artifacts may also occur due to inconsistencies between the
cameras.
For instance, \emph{exposure artifacts} ---i.e., very different brightness
between adjacent cameras--- may appear, and the lack of synchronization among
the cameras may result in \emph{motion discontinuities}.
If those issues are not handled properly by the video stitching step, they
will ultimately impact the overall pipeline and be perceived by the end user.

Omnidirectional stereoscopic 3D content capture is subject to typical
distortions of standard stereoscopic 3D content, such as \emph{keystone
distortion}, \emph{depth field curvature}, and \emph{cardboard effect}
~\cite{boev2008classification,hanhart20133d}.
\emph{Keystone distortion} is the result of the position of the two cameras
(for left and right eyes) converging to slightly different planes, which
causes a vertical parallax, i.e., a vertical difference between homologous
points~\cite{hanhart20133d}.
The same principle in the horizontal direction leads to the \emph{depth plane
curvature} artifact.
The \emph{cardboard effect} refers to an unnatural flattening of objects in
stereoscopic images--affected objects appear as if they were cardboard cut
outs~\cite{boev2008classification}.

Moreover, compared to capturing stereoscopic 3D content for cinema and TV,
capturing omnidirectional stereoscopic 3D content adds up their own set of
challenges~\cite{tan2018360-degree}.
For instance, the optical centers of individual cameras do not share the same
center of projection.
However, applying planar transformation models to synthesize multiple views
together on a common virtual surface is only valid if the captured scene is a
planar surface itself, or if the cameras share the same center of projection
~\cite{knorr2017modular}.
For off-centered cameras, transformation error increases with the off-center
distance and the amount of depth within the captured scene.
The warping and stitching process can finally worsen \emph{keystone} and
\emph{depth field curvature} issues.

\subsection{Stitching issues}

\begin{figure*}[!htb]
 \centering
  \begin{subfigure}[b]{.19\textwidth}
    \includegraphics[width=\textwidth]{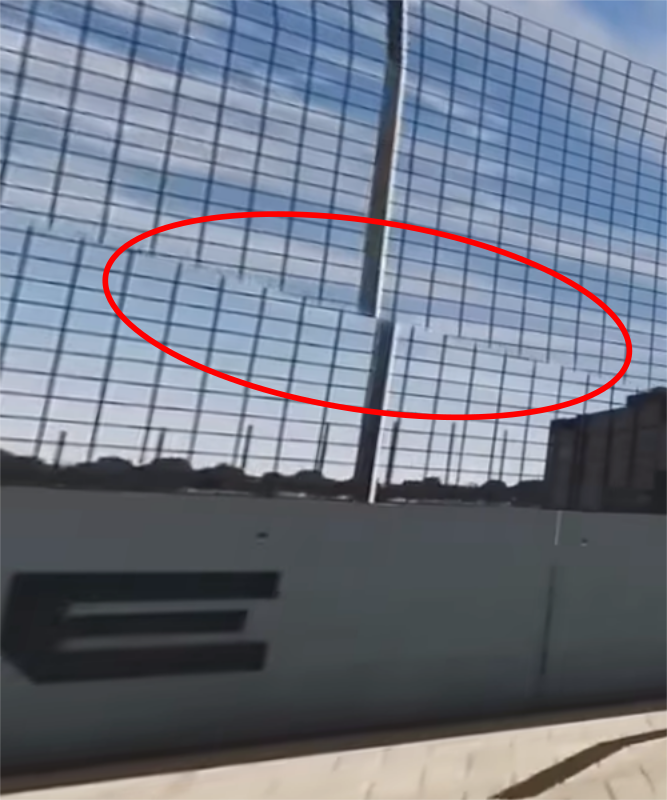}
    \caption{}
    \label{fig:stitching_discontinuities}
  \end{subfigure}
  \begin{subfigure}[b]{.19\textwidth}
    \includegraphics[width=\textwidth]{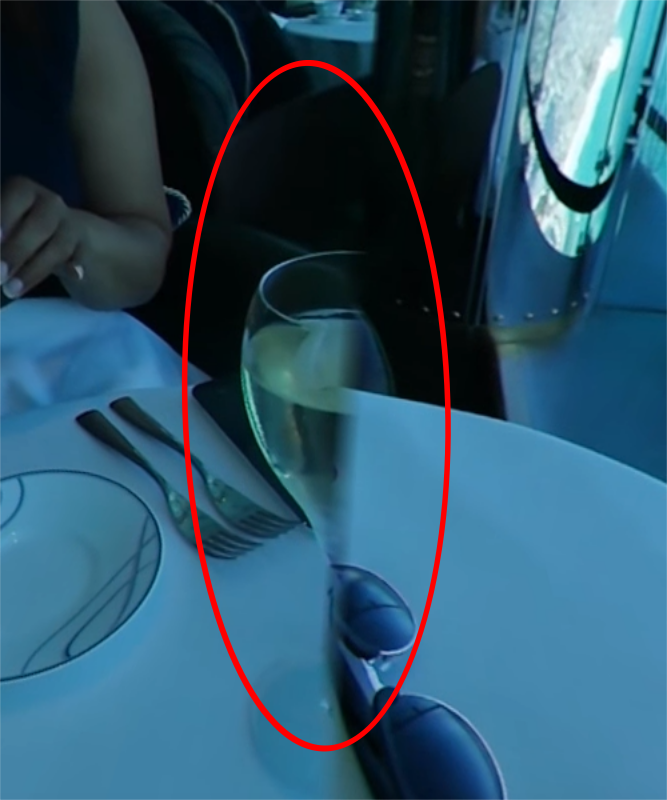}
    \caption{}
    \label{fig:stitching_missing}
  \end{subfigure}
  \begin{subfigure}[b]{.19\textwidth}
    \includegraphics[width=\textwidth]{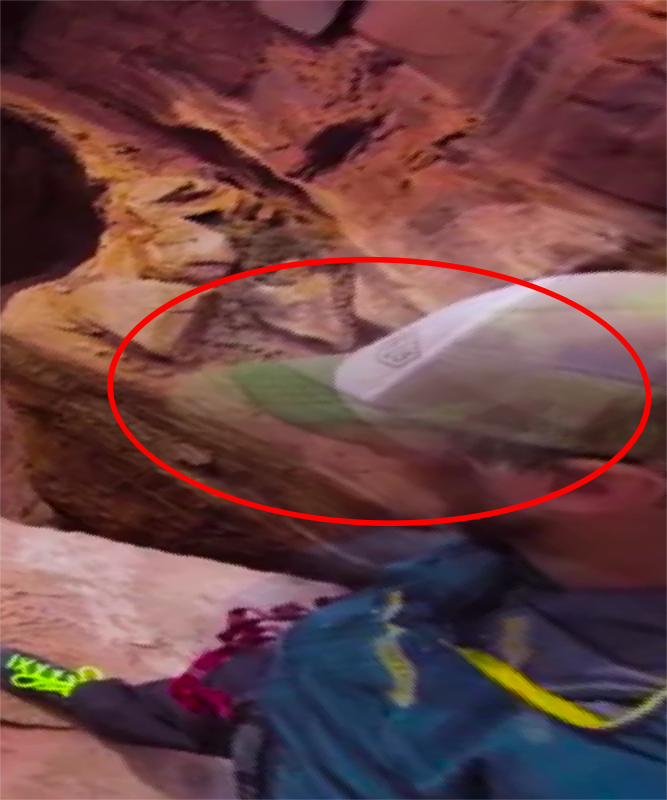}
    \caption{}
    \label{fig:stitching_ghosting}
  \end{subfigure}
  \begin{subfigure}[b]{.19\textwidth}
    \includegraphics[width=\textwidth]{imgs/artifacts/stitching_4.png}
    \caption{}
    \label{fig:stitching_geometrical_distortions}
  \end{subfigure}
  \begin{subfigure}[b]{.19\textwidth}
    \includegraphics[width=\textwidth]{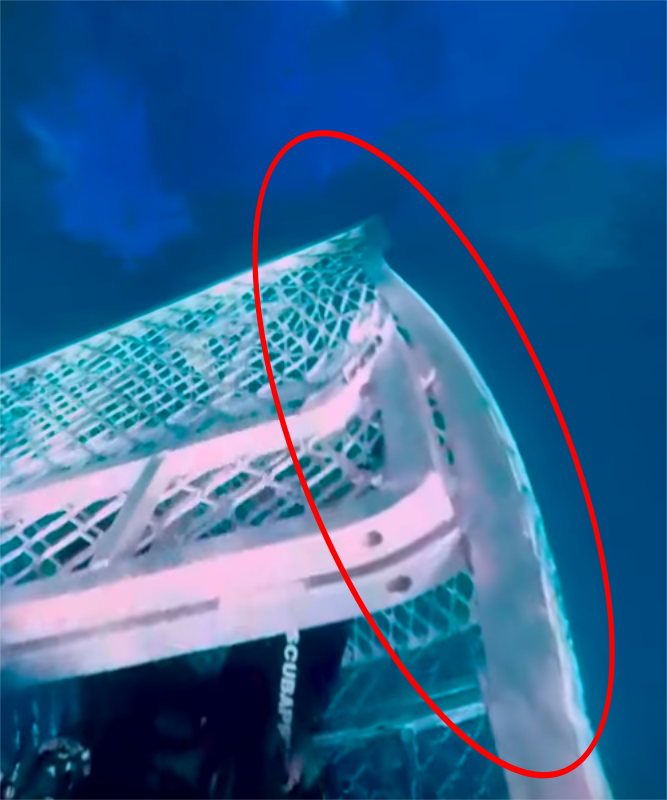}
    \caption{}
    \label{fig:stitching_geometrical_distortions}
  \end{subfigure}

  \caption{Examples of \emph{stitching} artifacts: (a)~broken edges and
           (b)~missing information;
           \emph{blending} artifacts: (c)~ghosting and (d)~exposure; and
           \emph{warping} artifacts: (e)~geometrical distortions / object
           deformations.}
  \label{fig:stitching_artifacts}
\end{figure*}

The unreliable information due to optical distortions and motion
discontinuities between the different cameras usually makes the stitching
process challenging.
(Indeed, given its complexity image and video stitching has been an active
research area~\cite{szeliski2007image}.)
Besides combining and warping the individual images to create the spherical
signal, the video stitching process may also have to compensate for some of
the sensor limitations and inconsistencies among the cameras in the
multicamera rig.
Due to these challenges, most approaches are still affected by some visually
annoying artifacts, which may appear as \emph{blurring}, \emph{visible
seams}~(due to different exposures to color and brightness discontinuities),
\emph{ghosting}, \emph{misaligned/broken edges and image structures},
\emph{missing information}~(e.g., objects with missing parts), and
\emph{geometrical distortions}~(e.g., visible deformation of objects).
Fig.~\ref{fig:stitching_artifacts} exemplifies some of these artifacts.
It is important to note that some stitching algorithms estimate depth
information by multiview geometry estimation, in order to have a better
reconstruction.
As a result, objects close to the cameras tend to be more affected by
stitching errors, because the depth errors are more significant in these
areas.

\begin{figure}[!htb]
  \centering
  \begin{subfigure}[b]{.49\columnwidth}
    \includegraphics[width=\columnwidth]{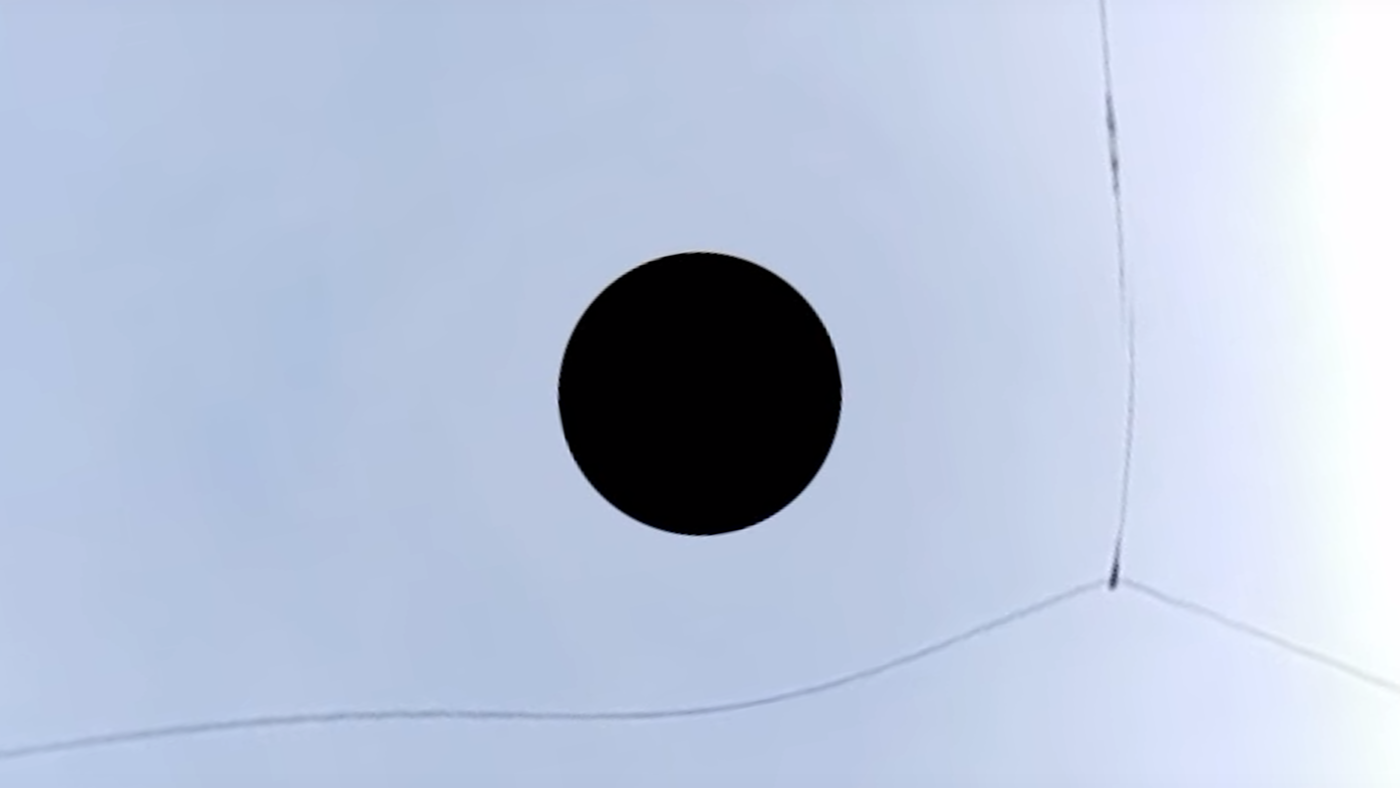}
    \caption{}
  \end{subfigure}
  \begin{subfigure}[b]{.49\columnwidth}
    \includegraphics[width=\columnwidth]{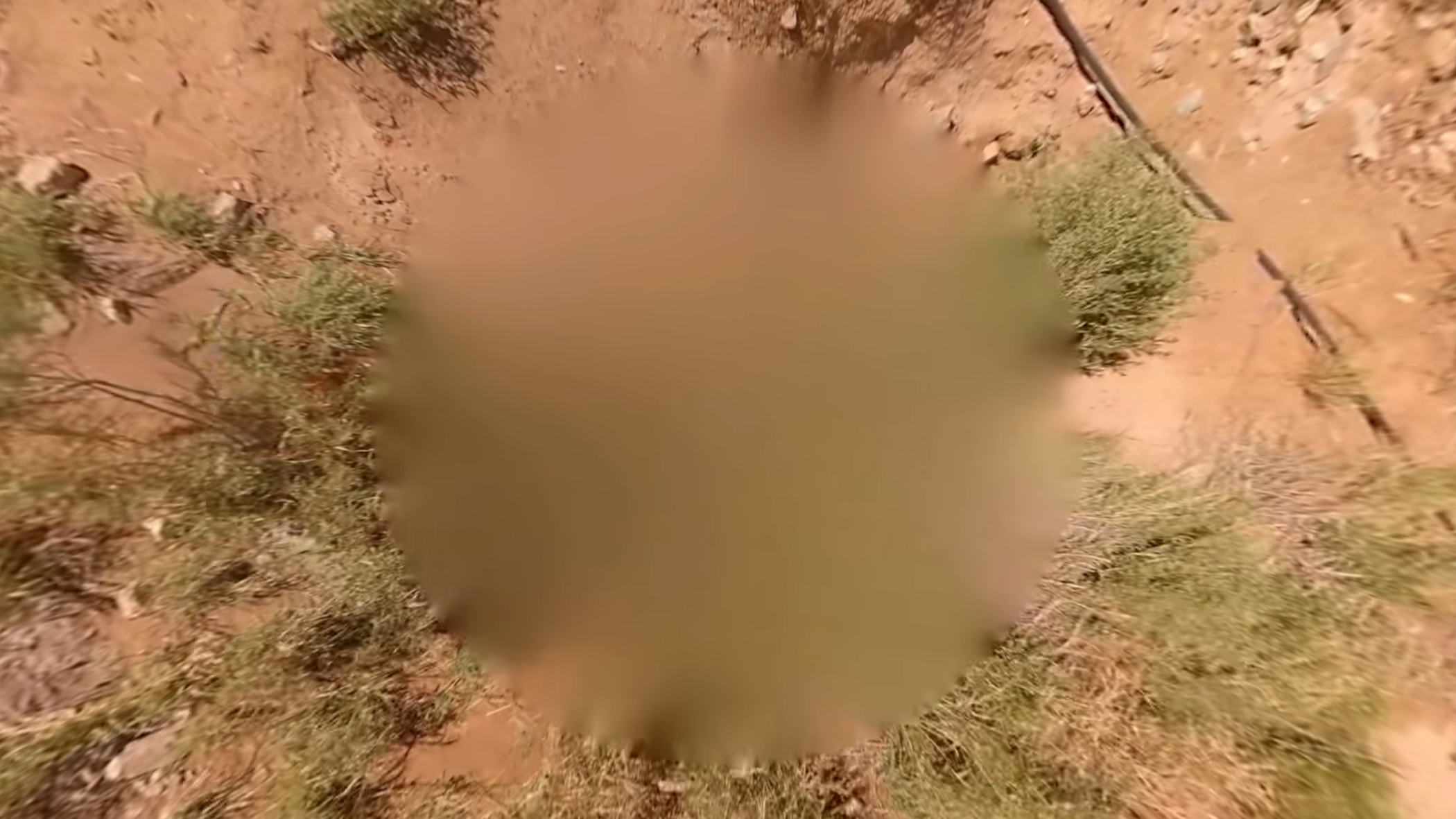}
    \caption{}
  \end{subfigure}
  \caption{Examples of post-processing on the poles due to missing areas:
           (a)~a black circle and (b)~a (inpainted) blurred circle.}
  \label{fig:missing_area}
\end{figure}

Another artifact that may be present in some 360-degree content is the
\emph{black circle} or \emph{blurred circular areas} on
poles~(see~Fig.~\ref{fig:missing_area}).
Such an artifact is mainly due to: the multicamera rig being unable to capture
the full 360-degree field of view; or post-processing to remove the camera
stand of the scene through inpainting techniques.
Note that it is also quite common to replace this pole area with the name of
the camera brand or alternative information.

In video settings, it may further be possible to see \emph{motion
discontinuities}, such as \emph{object parts appearing and disappearing}
abruptly and \emph{temporal geometrical distortions}, when moving objects come
across the stitching areas.
Moreover, for some camera rigs, if the video stitching algorithm does not
successfully treat the lack of synchronization among individual cameras,
\emph{synchronization issues} may be perceived by the end user. 
For instance, temporal inconsistent stitching of the camera views can result
in unsteady scene appearance over time on the stitching areas, resulting in
\emph{wobbling artifacts}.
\ranote{What about abruptly changes of seams? See Qualcomm Virtual Reality
Content Creation Technology report.}

Interestingly, the regions affected by stitching artifacts are generally known
for a given camera rig.
Thus, from a cinematic point of view, it is usually a good practice to have
less action and useful information in this area\ranote{~\cite{?}}.

Finally, compared to capturing monoscopic 360-degree video, capturing
omnidirectional stereoscopic video usually increases the amount of stitching
and blending errors.
On the one hand, to reduce stitching errors, the baseline between the cameras
should be minimized.
On the other hand, the baseline between the cameras of different views needs
to be increased in S3D content creation as parallax is required for generating
a 3D effect~\cite{knorr2017modular}.
In addition, minor flaws in the footage or stitching errors are usually
magnified when viewed in stereoscopic 3D. \nbnote{\bf{WHY? AND/OR
REFERENCES?}}
Given that such errors can occur in different places in each view, this can
result in \emph{binocular rivalry} and \emph{discomfort} when watching
stereoscopic 360-degree content.




%
%
%
%

\section{Artifacts caused by Encoding}
\label{sec:coding}

\subsection{Projection to the coding geometry}
\label{subsec:projection}

\begin{figure}
  \centering
  \begin{subfigure}[b]{.9\columnwidth}
    \includegraphics[width=\columnwidth]{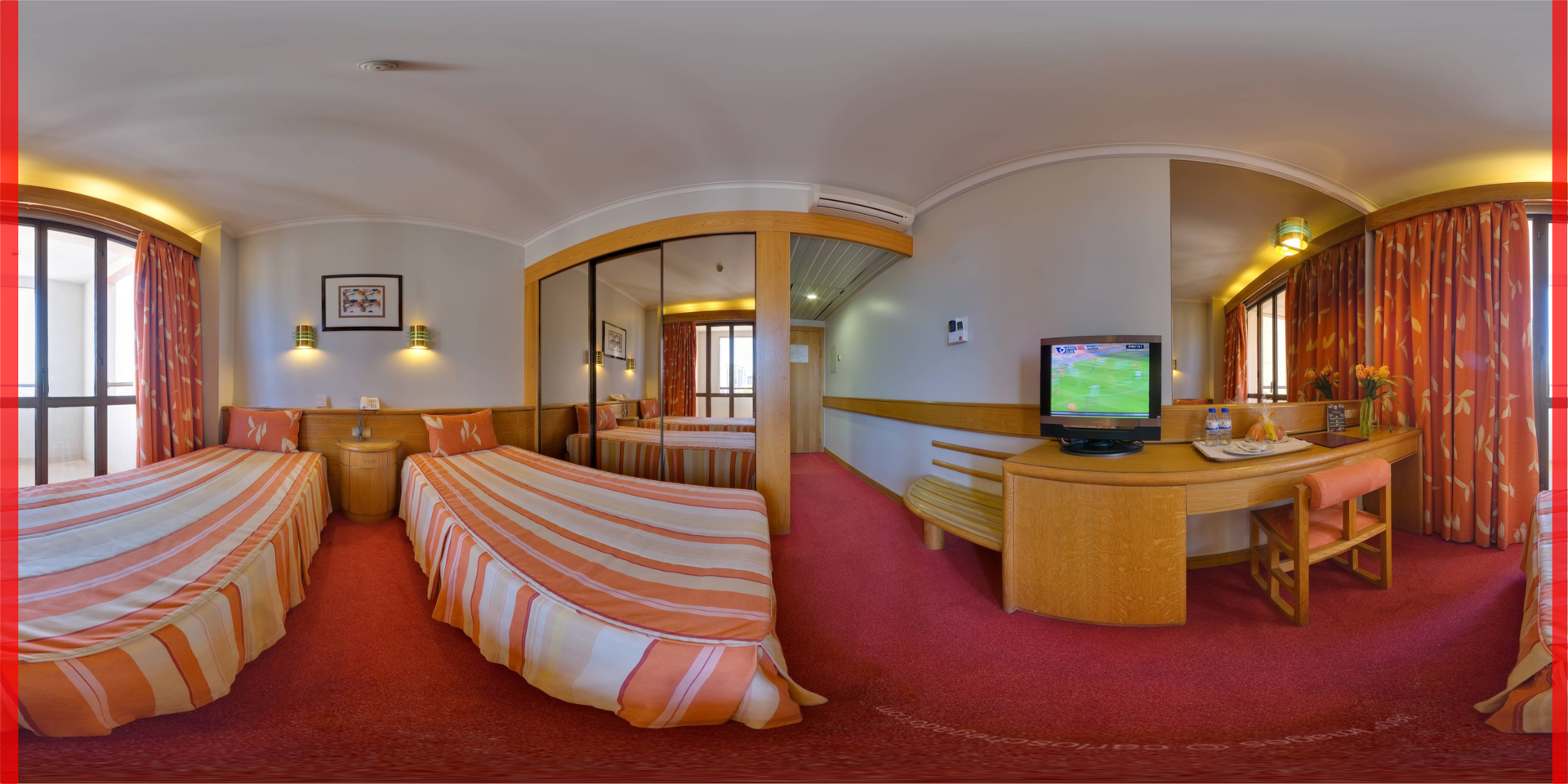}
    \caption{}
  \end{subfigure}
  \begin{subfigure}[b]{.9\columnwidth}
    \includegraphics[width=\columnwidth]{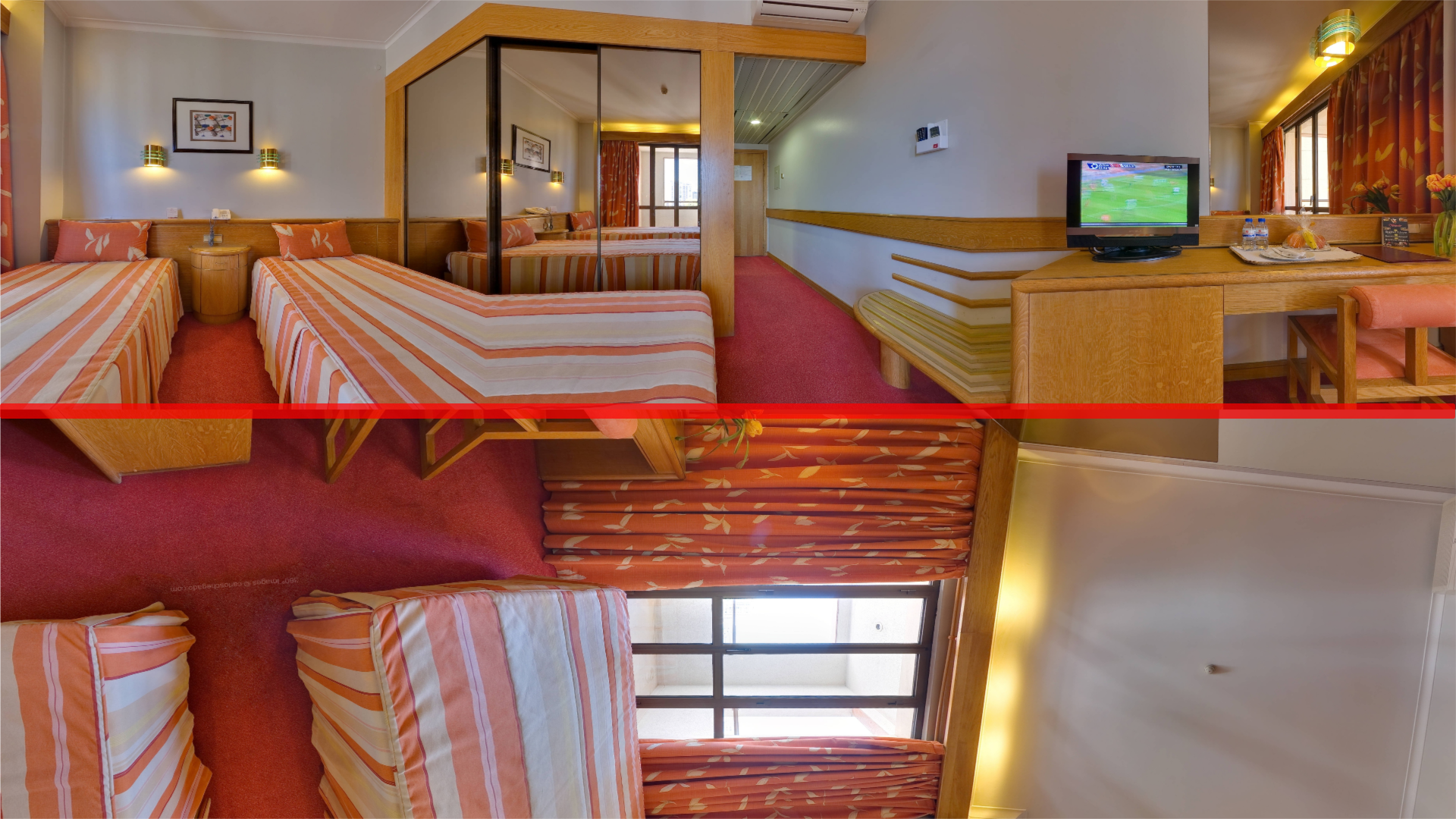}
    \caption{}
  \end{subfigure}
  \caption{Examples of discontinuities implied by the (a)~ERP and (b)~CMP
           representations.
           In~(a) the vertical borders of the ERP representation are not
           neighbors on the planar representation, but they are neighbors on
           the spherical domain.
           In~(b) discontinuities also happen on the borders of the frame,
           but, in addition, due to \emph{frame packing} some faces which are
           not neighbors on the spherical domain, became neighbors on the
           planar representation.}
  \label{fig:discontinuities_planar_domain}
\end{figure}

Projecting a sphere to a plane is a common problem in map
projections~\cite{pearson1990map}, and it is impossible to do so without
adding some \emph{geometrical distortions} and \emph{discontinuities}~(i.e.,
neighboring regions on the spherical domain may end up not being neighbors on
the planar representation, or vice-versa) on the planar representation.
Different projections may imply different \emph{geometrical distortions} and
\emph{discontinuities} regions.
Fig.~\ref{fig:discontinuities_planar_domain} shows examples of geometrical
distortions and discontinuities resulted from the ERP and CMP map projections.
Even though these distortions are not directly viewed by the end user, the
interaction between the geometrical distortions and the lossy compression
processing may result in visible artifacts.

\begin{figure}[!htb]
  \centering
  \begin{subfigure}[b]{0.9\columnwidth}
    \includegraphics[width=\columnwidth]{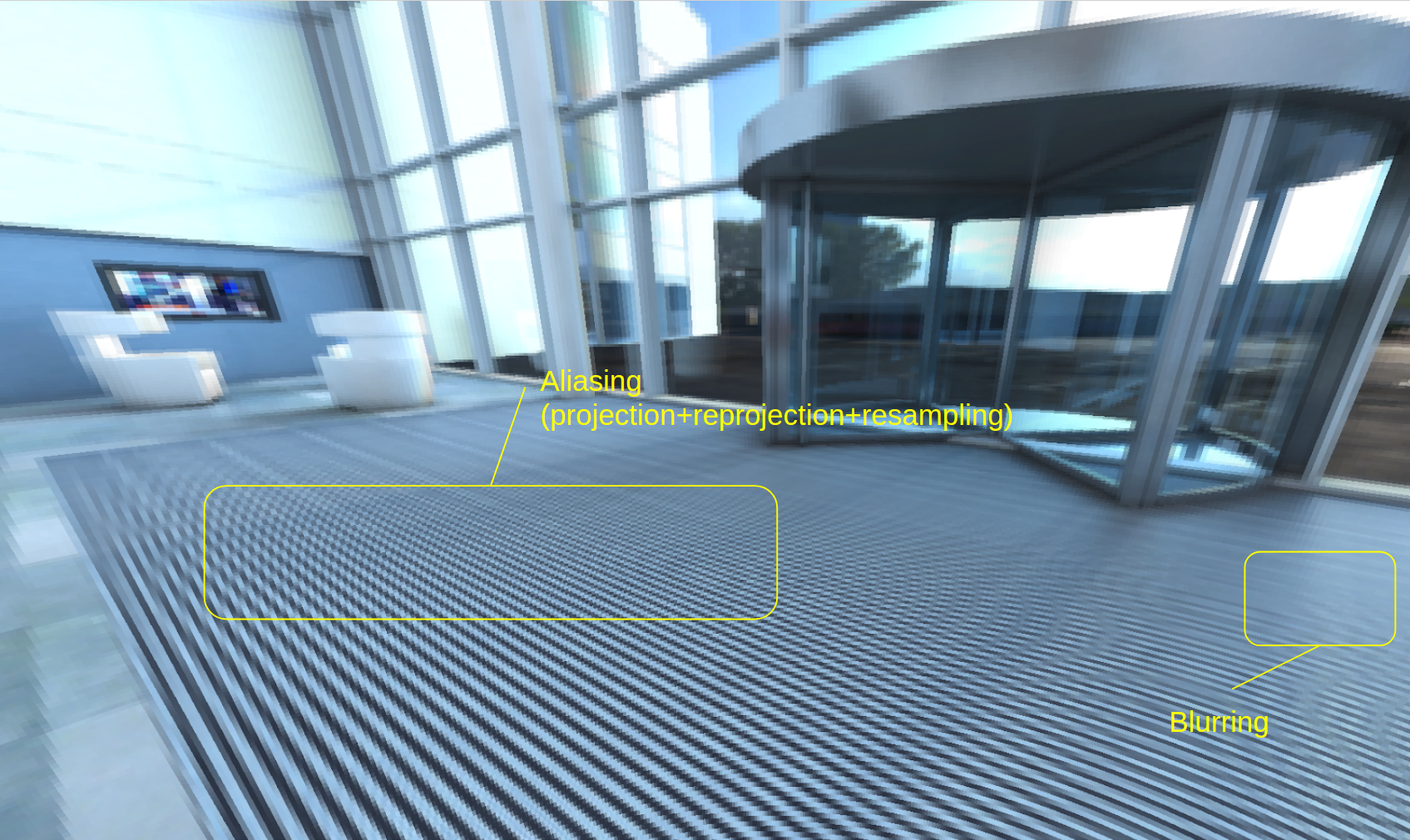}
    \caption{}
    \label{fig:projection_aliasing_blurring}
  \end{subfigure}

  \begin{subfigure}[b]{0.9\columnwidth}
    \includegraphics[width=\columnwidth]{imgs/artifacts/visible_pole_sampling.png}
    \caption{}
    \label{fig:projection_visible_pole}
  \end{subfigure}

  \begin{subfigure}[b]{0.9\columnwidth}
    \includegraphics[width=\columnwidth]{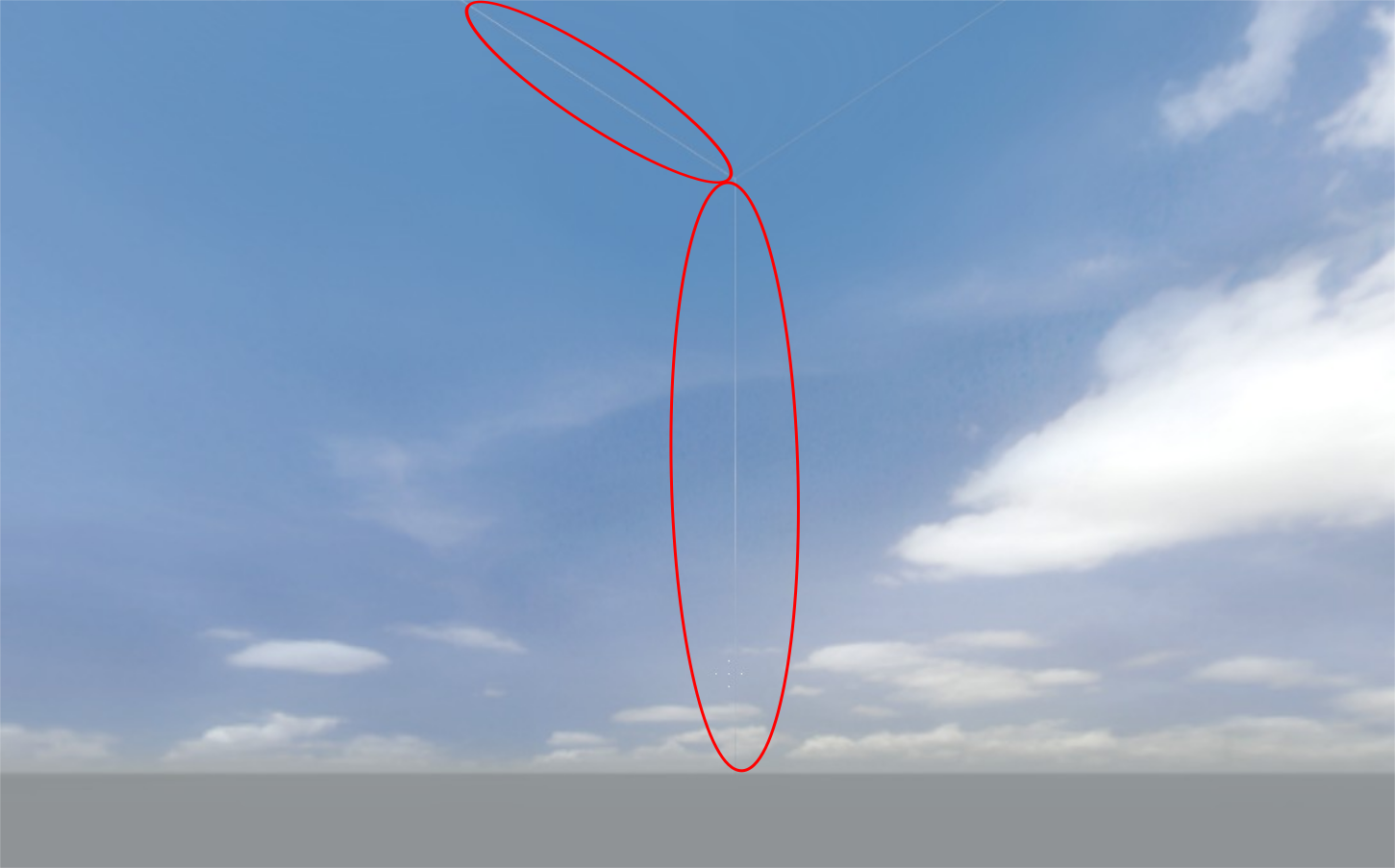}
    \caption{}
    \label{fig:projection_seams_sampling}
  \end{subfigure}

  \caption{Examples of (a)~aliasing and blurring, (b)~visible poles; and
           (c)~visible seams due to projection and resampling.} 
  \label{fig:projection_artifacts}
\end{figure}

Moreover, since the projection from the spherical to the planar representation
(and the back-projection to the spherical domain at the client side) involves
some resampling and interpolation, different map projections may result in
\emph{aliasing}, \emph{blurring}, and \emph{ringing} distortions in the signal
visualized by the end-user~(see Fig.~\ref{fig:projection_aliasing_blurring}).

Also, if sampling and interpolation are not treated correctly additional
distortions may happen, such as: \emph{visible poles} due to oversampling on
the poles areas, may appear when using the ERP representation~(see
Fig.~\ref{fig:projection_visible_pole}); and \emph{visible seams} in the
discontinuities regions~(see Fig.~\ref{fig:projection_seams_sampling}).
Methods like graph-based techniques~\cite{bagnato2012plenoptic} that are well
adapted to the specific geometry of images could reduce such artifacts by
processing the data in their native geometry.
However, current 360-degree systems exclusively rely on sampling and
interpolation techniques in the classical rectangular geometry.



%

\subsection{Compression}
\label{subsec:compression}

\begin{figure*}[!htb]
  \centering
  \begin{subfigure}[b]{0.5\textwidth}
    \includegraphics[width=\columnwidth]{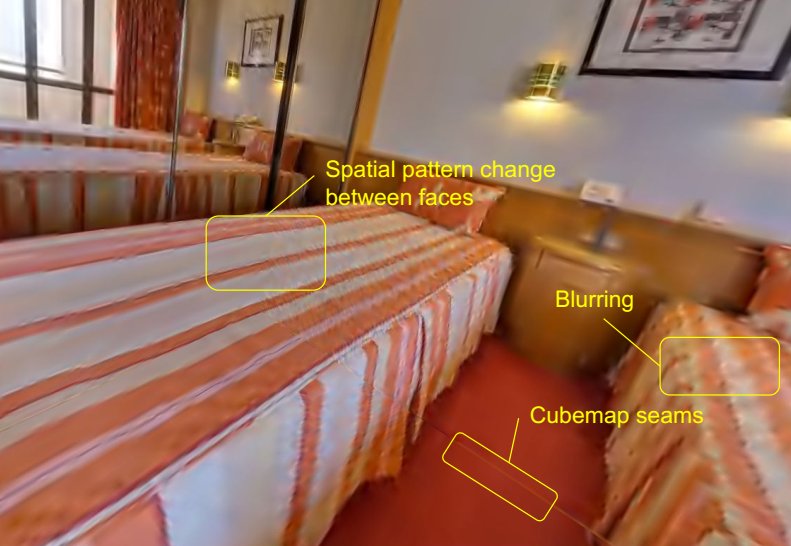}
    \caption{}
    \label{fig:cubemap_seams_example_jpeg2000}
  \end{subfigure}
  \begin{subfigure}[b]{0.475\columnwidth}
    \includegraphics[width=\textwidth]{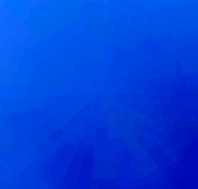}
    \caption{}
    \label{fig:compression_artifact_radial_blocking}
  \end{subfigure}
  \begin{subfigure}[b]{0.49\columnwidth}
    \includegraphics[width=\textwidth]{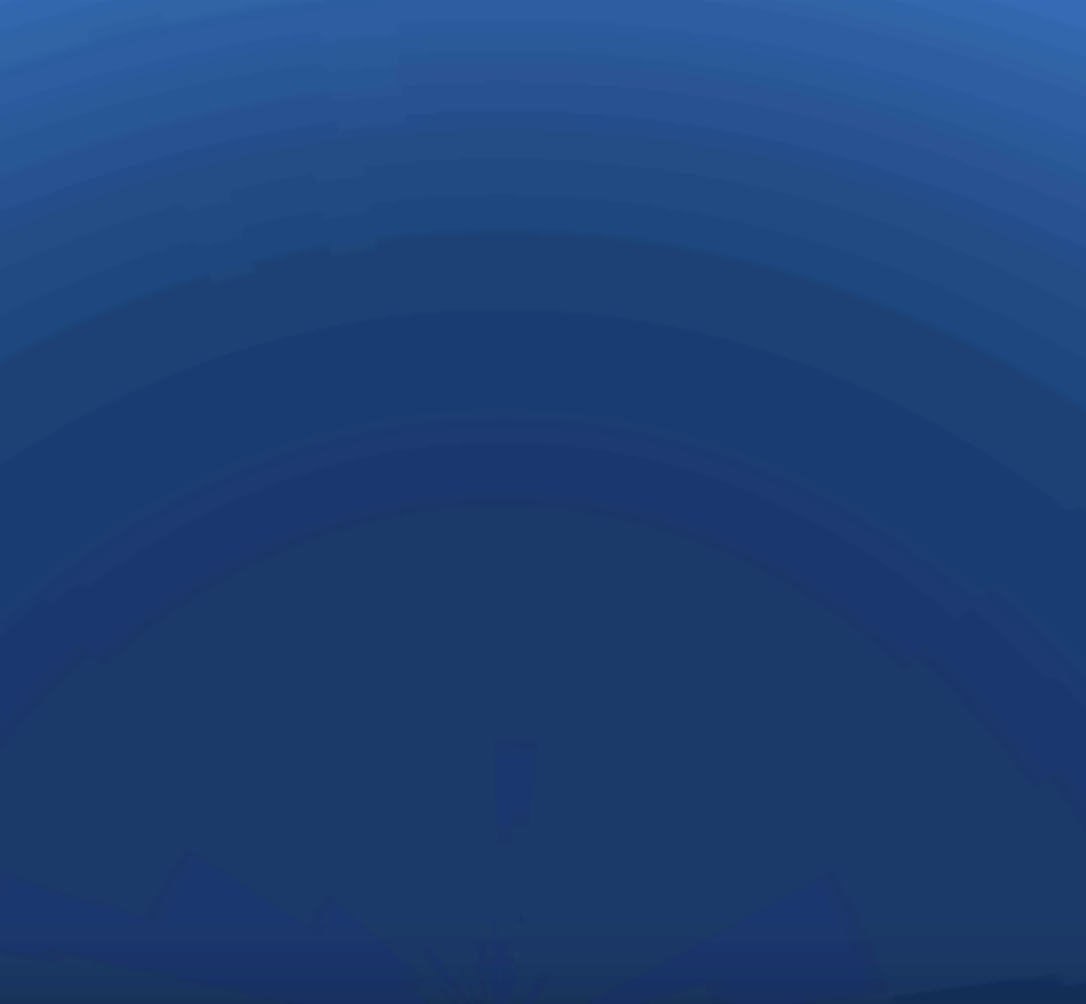}
    \caption{}
    \label{fig:compression_artifact_radial_banding}
  \end{subfigure}
  
    \begin{subfigure}[b]{.49\columnwidth}
    \includegraphics[width=\columnwidth]{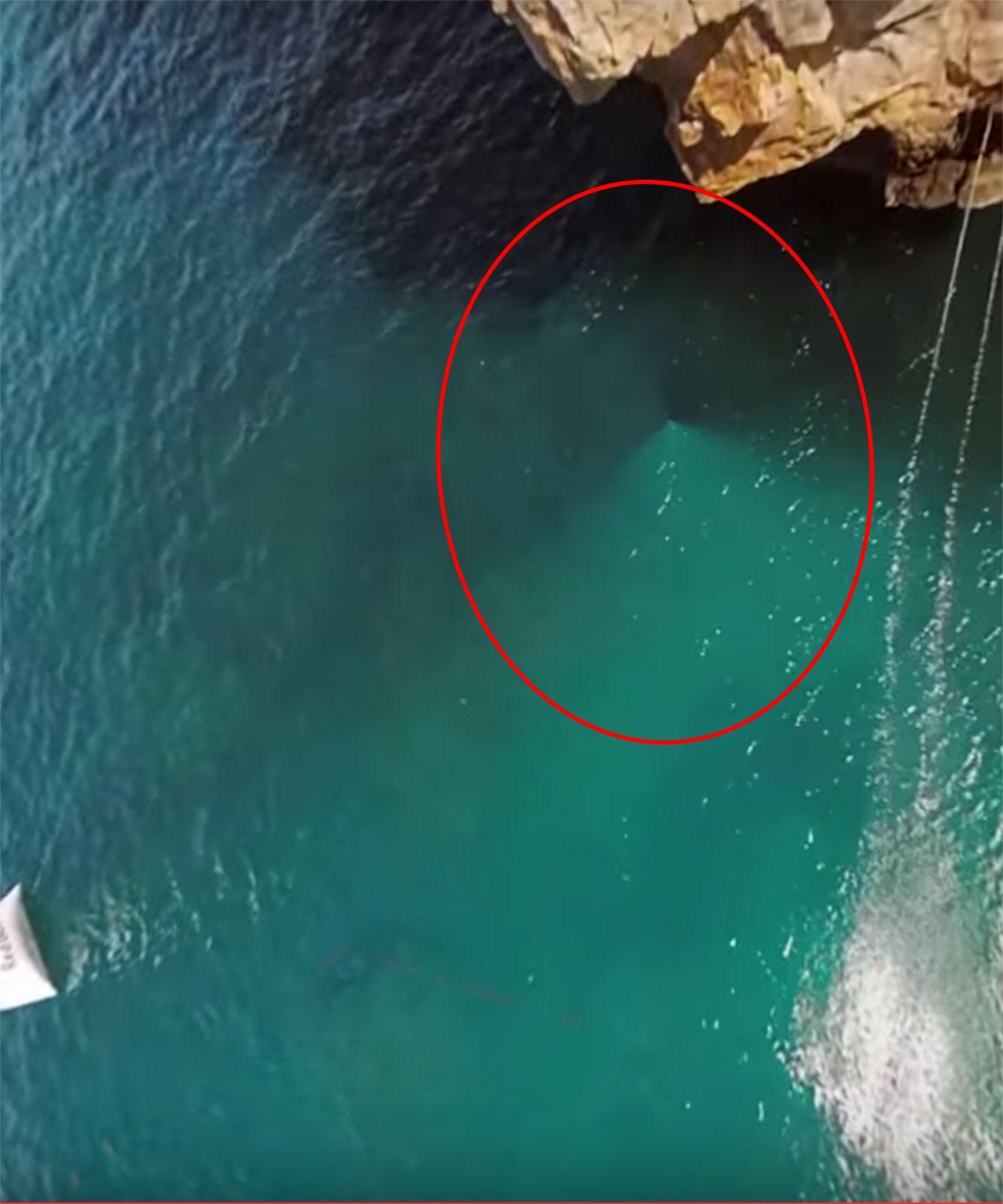}
    \caption{}
    \label{fig:visible_pole}
  \end{subfigure}
  \begin{subfigure}[b]{.5\columnwidth}
    \includegraphics[width=\columnwidth]{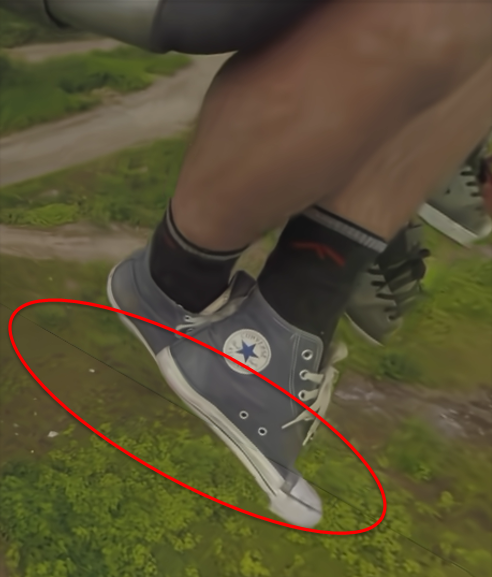}
    \caption{}
  \end{subfigure}
  \begin{subfigure}[b]{.415\columnwidth}
    \includegraphics[width=\columnwidth]{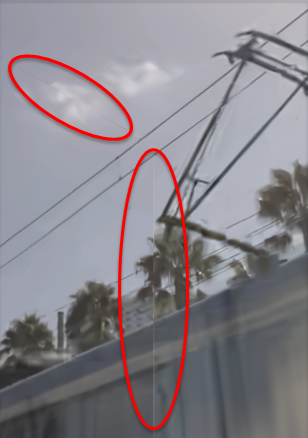}
    \caption{}
  \end{subfigure}
  \begin{subfigure}[b]{.492\columnwidth}
    \includegraphics[width=\columnwidth]{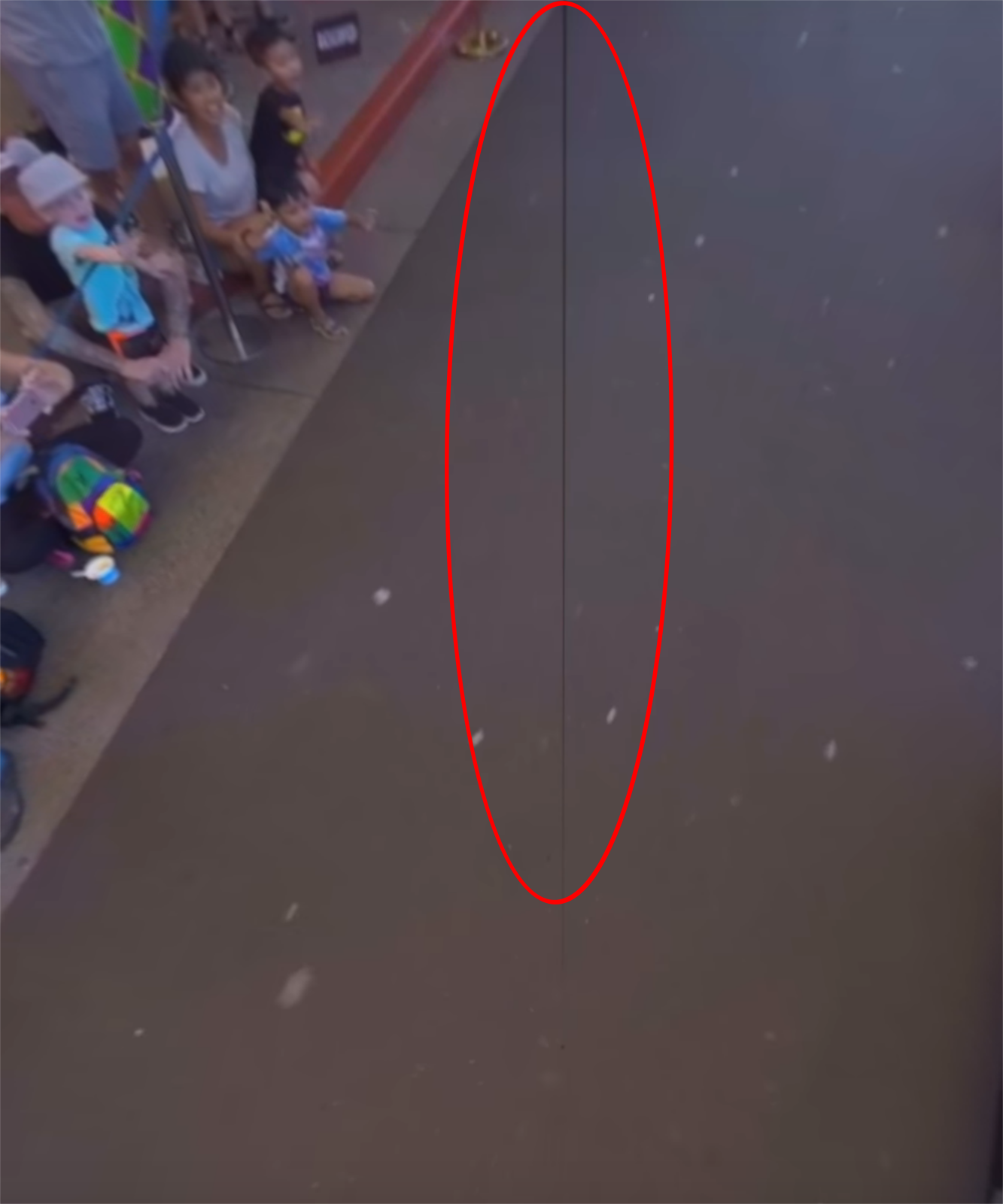}
    \caption{}
  \end{subfigure}

  \caption{Examples of compression artifacts on 360-degree content:
           (a)~blurring, cubemap seams, and spatial pattern changes (due to
               jpeg2000 compression);
           (b)~radial blocking;
           (c)~radial banding;
           (d)~visible pole;
           (e)-(f)~cubemap seams artifacts; and
           (g)~equirectangular seam artifact.}
  \label{fig:compression_artifacts}
\end{figure*}


Since the current approaches use conventional 2D video compression schemes for
the planar representation, they are also subject to the same artifacts
thoroughly studied in 2D video, which are briefly presented in
Table~\ref{tbl:2d_compression_artifacts}.
For a more in-depth discussion on 2D video artifacts, we refer the reader
to~\cite{zeng2014characterizing,unterweger2013compression,wu2005coding,yuen1998survey}.
Mostly, the origins of these artifacts in lossy block-based transform video
coding are~(directly or indirectly) due to quantization errors in the
transform domain~\cite{unterweger2013compression}.

\begin{table}[!h]
  \caption{Summary of conventional 2D video compression artifacts.
           \ranote{Remove this table if we need space.}}
  \label{tbl:2d_compression_artifacts}
  \begin{tabularx}{\columnwidth}{p{2cm} | p{6cm}}
    Artifact & Characteristics \\
    \hline
    \emph{Blocking} & is related to the appearance of the division of the
                      macroblocks; it is caused by coarse quantization of
                      low-detail regions.
                      \\
    \hline
    \emph{Blurring} & is the result of loss of spatial details in moderate-
                      high-detail regions; it occurs when high-frequency
                      components in the transform domain are quantized to zero
                      or due to strong deblocking filters.
                      \\
    \hline
    \emph{Color bleeding} & is the smearing of colors between areas of
                            strongly contrasting luminance; it happens due to
                            inconsistent image rendering on separately
                            compressed color channels or due to
                            interpolation on chroma-subsampled images/videos.
                            \\
    \hline
    \emph{Ringing} & appears as ``halos''~(artificial wave-like or ripple
                     structure) around sharp edges, e.g., strong edges and
                     lines.
                     \\
    \hline
    \shortstack{\emph{Stair case} and \\ \emph{basis pattern}} 
                    & incapability of horizontal and vertical basis
                      functions~(as building blocking of the DCT and its
                      variations) to accurately represent diagonal
                      edges~(similar to steep edges).
                      \\
    \hline
    \emph{Flickering} & refers to frequent changes in luminance or chrominance
                        along the temporal dimension that do not
                        appear in uncompressed video, and can be divided into
                        \emph{mosquito noise} (when it occurs at the borders
                        of moving objects), \emph{coarse-granularity
                        flickering}~(when it suddenly occurs in large spatial
                        areas) and \emph{fine-granularity flickering}~(when it
                        appears to be flashing on a frame-by-frame
                        basis)~\cite{zeng2014characterizing}.
                        \\
    \hline
    \emph{Jerkiness} & occurs when the temporal resolution is not high enough
                       to catch up with the speed of moving objects, and thus
                       the object motion appears to be discontinuous.
                       \\
    \hline
    \emph{Floating} & is the appearance of illusive motion in certain regions
                      as opposed to their surrounding background; the illusive
                      motion is erroneous because these regions are supposed
                      to stay or move together with the background.
                      \\
  \end{tabularx}
\end{table}

\ranote{Due to sampling in the projection step and luminance changes during
compression, visible poles can also appear~(see
Fig.~\ref{fig:visible_pole}).~\ranote{ERP-only?}}


%

%


%

With its particular geometry, the omnidirectional video is generally affected
by a complex combination of the compression artifacts that affect the
rectangular frames, as well as the frame packing and the warping due to the
map projection used.
For example, the blocking artifacts produced by the compression in the planar
domain will also be warped due to the omnnidirectional geometry.
Thus, they will be perceived as different \emph{warped blocking patterns},
which depend on the underlying geometry of the map projection.
For instance, in the ERP representation, blockiness close to the poles may be
perceived as a \emph{blocking radial pattern} by the end-user~(see
Fig.\ref{fig:compression_artifact_radial_blocking}).
Similarly, for the CMP representation, it may be possible to see the
\emph{perspective projection of the blocking artifacts}, and eventually
\emph{identify the underlying cube faces}.


%



As previously mentioned, inevitably, when using a 2D rectangular image to
represent the full 360-degree spherical signal, some neighboring regions in
the spherical domain are not neighboring in the planar representation.
Thus, when such a region is coded in the planar domain~(without taking into
account the original neighbors) and projected back to the spherical domain,
\emph{discontinuities} or \emph{visible seams} can appear.
For instance, when using ERP, a seam can appear in the region closing the
sphere, whereas unnatural seams can appear on some cube edges when a CMP
representation is used instead~(see
Fig.~\ref{fig:discontinuities_planar_domain}).
The origins of those visible seams due to compression can also be traced back
to:
(1)~transform blocks falling between two faces in the planar representation;
(2)~color bleeding or ringing artifacts from one face to the other;
and
(3)~deblocking filter algorithms that mismatch the faces discontinuities as
    blocking artifacts, and thus may smear content from one face to
    another~\cite{sauer2018geometry-corrected}.
In both cases some data from one face bleeds to the neighboring one in the
planar domain.
Since those two faces are not neighbors on the spherical domain, seams may
become visible.
Due to changes on the properties of the \emph{visible seams} during
consecutive frames, it is also possible to end up with \emph{flickering seams} 
in the temporal domain.


On coarse quantized lossy compressed video, the appearance of \emph{blocking},
\emph{blurring}, \emph{staircase} and \emph{basis pattern} in combination with
the warping and frame arrangement on the planar representation may also result
in visible \emph{spatial pattern transitions} on the viewports.
This is mainly the result of the different compression distortions being
applied in different directions in the planar domain, when compared to the
viewports.
This is the case, for instance, in CMP, where each different face may undergo
different geometrical distortions and rotations, which may cause visually
noticeable texture area changing its underlying ``pattern'' across adjacent
CMP faces~(see Fig.~\ref{fig:cubemap_seams_example_jpeg2000}).
In the temporal dimension, if an object is crossing from one face to the
other, it may alse be possible to see dynamic changes on its underlying
``pattern''.

Moreover, the use of compressors unaware of the geometry of omnidirectional
videos also result in the content being more prone to motion compensation and
flickering issues than the classical video counterparts.
Most modern video codecs use block-based motion estimation for inter
frame compression ---i.e., a block of pixels is matched to neighboring frames
(and usually from blocks on neighboring areas, to speed things up), and if
there is a good match, a 2D offset vector~(smaller than a block) is
calculated and stored instead of the original block.
Indeed, if blocks are small enough, block vectors can represent general planar
motion with perspective cameras rotating in all 3 axes.
However, the planar representation of the 360-degree content implies that at
some parts the motion is no longer planar and vectors cannot be predicted so
well from neighbours.
Thus, the motion model and intra prediction is not optimal in regions such as
the poles on ERP and in discontinuities in CMP, which may result in higher
bitrates and compression artifacts, such as motion flickering, in these
areas.~\cite{desimone2017deformable}

\ranote{As for the acquisition artifacts, we note that....Due to the
        interaction between projection, resampling, and compression,
        high-textured areas \emph{flickering} are more probable in 360-degree.
        YES!!!}

Finally, in stereoscopic settings, the compressed ODS video is also subject to
the same artifacts that have been studied in the context of stereoscopic 3D
video~\cite{boev2008classification,hanhart20133d}.
One of the leading sources of compression-related stereoscopic artifacts is
the possibility of the compression algorithm to introduce different
distortions on the left and right frames, resulting in \emph{cross
distortion}, which may affect depth perception and cause binocular
rivalry.
Visible seams artifacts, which are specific of 360-degree content,  may also
be affected by cross-distortions, and may result in a volumetric perception of
the seams.
When using a frame-compatible approach for the ODS content, the
discontinuities between the left and right content may also result in the
appearance of new seams
\ranote{Finally, distortion effects are amplified in omnidirectional videos,
by a combination of...}
Asymmetric stereoscopic 3D spatial resolution and
compression~\cite{naik2018optimized}
is also another potential source for cross distortions.
The \emph{cardboard effect} may also be introduced by compression.
\ranote{In addition to the above artifacts, omnidirectional stereoscopic video
        suffers from distortions due to the specific geometry of the images,
        In particular....}

\section{Transmission-related artifacts}
\label{sec:transmission}

Transmission delays and communication losses affect the streaming of
omnidirectional video sequences, similarly to how they affect traditional videos.
As the recent streaming systems are based on adaptive streaming algorithms, we
focus on their specific artifacts in the following.

Depending on the adaptive streaming scheme
~(\emph{viewport-agnostic}, \emph{viewport-dependent projection}, or
\emph{tile-based}) in use and on the implemented adaptation logic for the
360-degree content, different distortions can appear and impact the user
experience.

In the \emph{viewport-agnostic} adaptive streaming, the typical DASH
distortions, such as \emph{delay}, \emph{rebuffering events} and \emph{quality
fluctuation}, may be perceived in the user's field of view.
These distortions have been widely studied and characterized for conventional
2D video content and displays~\cite{Garcia2014, Seufert2015} and few studies
have investigated them in the context of stereoscopic 3D
video~\cite{hamza2014dash}.
The impact of these distortions on the Quality of Experience~(QoE) of
immersive applications when the compressed content is projected to the
viewport and viewed through an HMD is still largely overlooked~\cite{Schatz2017}.

In the \emph{viewport-aware} adaptive streaming schemes, the ability of the
system to predict how the user navigates the content can also impact the
artifacts perceptible by the end user. Besides the typical DASH-based
distortions, new artifacts may appear.

In the \emph{viewport-dependent projection} approach, besides the temporal
quality fluctuations in the field of view, there is also the possibility of
the user experiencing \emph{quality fluctuations} and \emph{rebuffering
events} \emph{on head movement}. Also, when the viewport is composed of regions
with different qualities, \emph{spatial quality fluctuations in the
viewport} may become annoying.

The \emph{tile-based} approach, depending on the adopted tiling scheme, is
also subject to \emph{spatial qualities fluctuations}~(when the viewport is
composed by tiles of different qualities).
In addition, the \emph{tile borders} may become visible and, in extreme cases,
a portion of the viewport could be missing~(\emph{incomplete viewport}) if the
client has not downloaded the corresponding tile.

\begin{figure}
  \centering
  \includegraphics[width=.95\columnwidth]{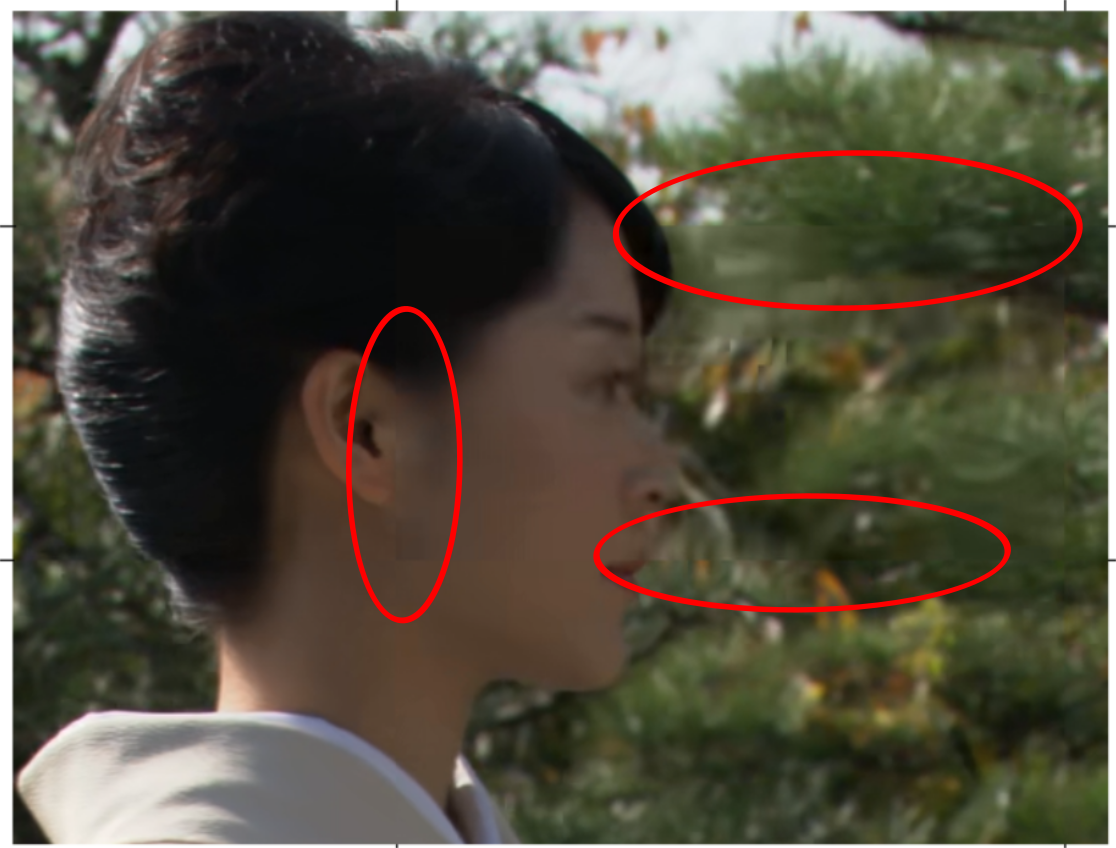}
  \caption{Example of \emph{spatial quality fluctuation} artifacts due to
           tiling~(highlighting the tile
           borders).~(Adapted from~\cite{concolato2017adaptive})}
\end{figure}

From the above visual artifacts, \emph{spatial quality fluctuations} in both
viewport-based adaptive streaming can also impact the two views differently,
leading to stereoscopic 3D artifacts.





\section{Artifacts from Displays}
\label{sec:display}

Even with a perfectly captured, transmitted, and received mono or stereo
omnidirectional image, artifacts still can appear due to technical limitations
of the current displays.
Among the common 360-degree visualization techniques, the HMD mode is the most
challenging one.
Indeed, all the artifacts of traditional displays, such as \emph{aliasing},
\emph{blurring}, \emph{motion blur}, etc.\ranote{\cite{?}}, may also affect
HMD displays.
In addition, new distortions that are specific to HMDs can appear due to the
fact that, compared with traditional displays, the HMD is very close to the
users eyes, it has a wider field of view, and, more importantly, it physically
moves with the user's head.
Such an interaction between a user's and display movement is unique to HMDs,
and it can cause new artifacts that have not been considered in traditional
displays, and that can even break the sense of presence or, worse, they can
make the user physically uncomfortable.

\begin{figure}[!htb]
 \centering
 \includegraphics[width=.6\columnwidth]{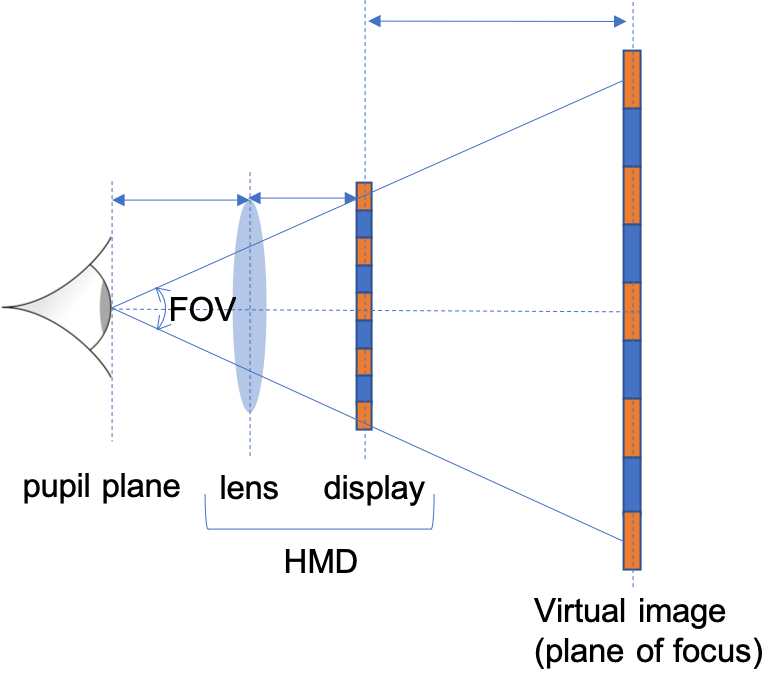}
 \caption{Simplified schematics of HMD}
 \label{fig:hmd_optics}
\end{figure}

Designed for supporting an immersive visual experience, most HMDs are composed
of a display device attached to the head~(and providing stereoscopic vision)
and an optical and a head-tracking system.
The purpose of the optical system~(see Fig.~\ref{fig:hmd_optics}) is twofold. 
First, the close distance between the user's eyes and the HMDs requires an
optical system to support comfortable content viewing.
Second, it serves to optically magnify the content presented on the screen,
supporting a Field of View~(FoV) closer to the natural human viewing.
The head tracking system allows the system to update the content presented to
the user based on his head position.

Different optical design systems have been used with the goal of supporting a
larger FoV and comfortable viewing on HMDs.
These systems have varying trade-offs in weight, field of view, light
transmittance, and image quality~\cite{patterson2006perceptual}, but they all
suffer from \emph{optical distortions}.
For instance, some lenses can cause \emph{chromatic aberrations} at the edge
of the FoV.
Currently, higher-quality HMD displays, such as Oculus and HTC, have changed
the design to incorporate fresnel lens~\cite{geng2018viewing} \ranote{Confirm
the ref.} features.
Although these new lenses improve on the rectification of the \emph{chromatic
aberrations}, they bring another problem sometimes referred to as
\emph{``god-rays''} or \emph{``flare''}, which is characterized by the
appearance of a halo at the FoV's edges.
This is mainly due to the light that is falsely redirected through the fresnel
steps.

On both of the aforementioned lenses types, the magnification characteristics
of HMDs is done by applying a significant \emph{pincushion} distortion through
the lenses.
Such a distortion must be rectified by applying a distortion in the other
direction, usually a barrel-distortion shader toward the end of the rendering
process.
The required amount of distortion is display specific, and if it is not done
properly it may also result in a \emph{barrel} or \emph{pincushion} distortion
perceived by the end user.
In both \emph{chromatic aberrations} and \emph{geometrical distortion},
shaders can be used to try to mitigate the visible
effects.~\cite{pohl2013improved}

\begin{figure}[!htb]
 \centering
 \includegraphics[width=\columnwidth]{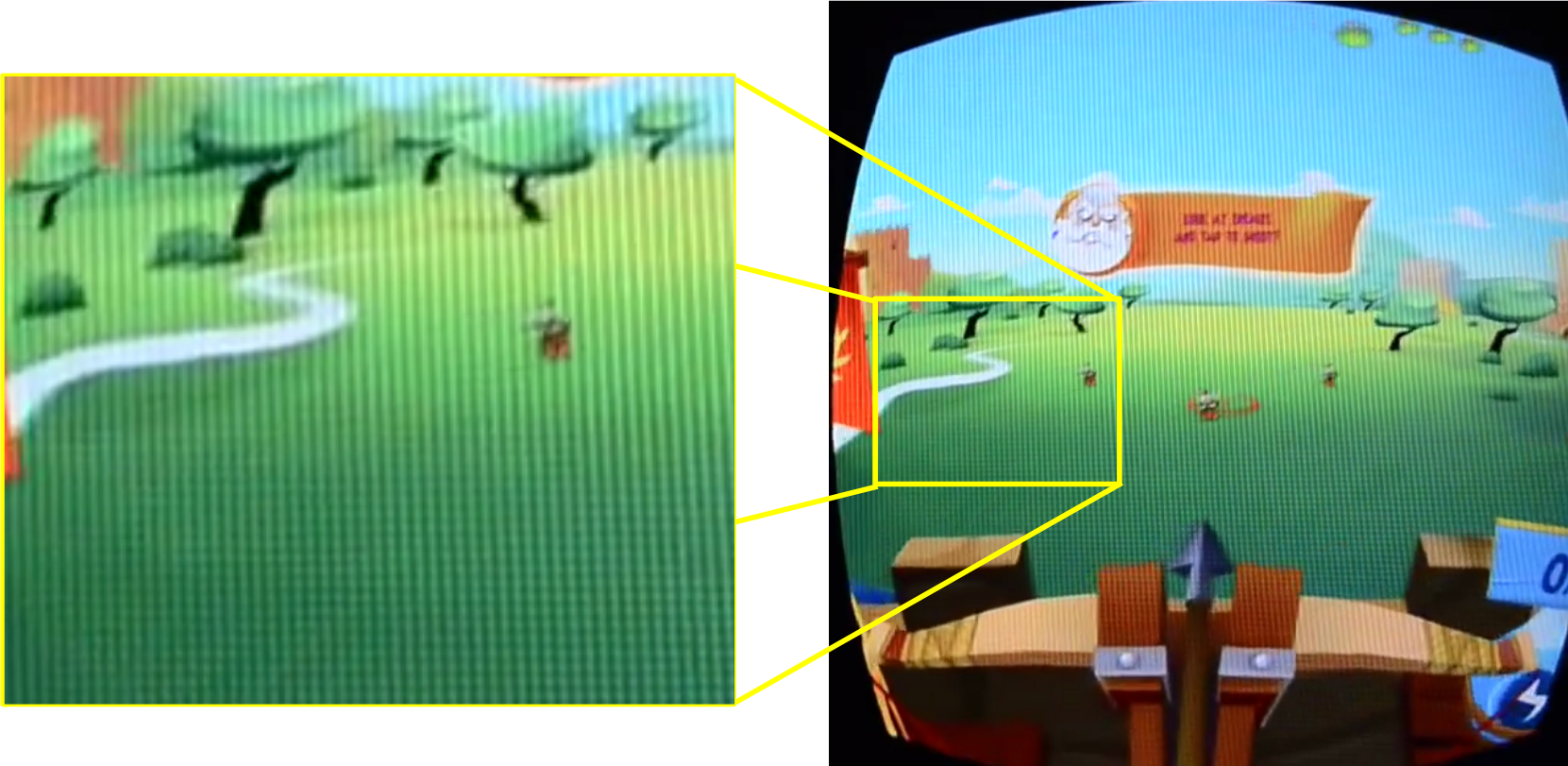}
 \caption[Example of the screen-door effect]{Example of the screen-door
    effect.~\footnotemark}
\label{fig:screen_door_effect}
\end{figure}

Then, when watching 360-degree video on most of the current HMDs, it is
possible to see a fixed lattice pattern~(such as the one shown on
Fig.~\ref{fig:screen_door_effect}) named the \emph{screen-door effect}.
Such a pattern mainly occurs because having the screen very close to viewers
eyes as in a HMD, it is actually possible to see the spacing between the
pixels.
The screen-door effect is certainly not a new phenomenon, but has been mostly
solved for the viewing distance of current digital TVs and projectors. 
For current HMD displays this is still an issue, and it may be solved in the
coming years with higher resolution displays.

\footnotetext{Frame extracted from \url{https://youtu.be/V7uTnOYLhZA}}



\emph{Motion-to-photon} delay is another artifact that is specific to HMDs. It
is defined as the time perceived by the end-user between his movement and the
full response on the display screen~\cite{zhao2017estimating}.
Despite being an annoying artifact, motion-to-photon delays may also induce
motion sickness.
Ideally, to achieve a full sense of presence, no motion-to-photon delay should
be perceived.



\begin{figure}[!htb]
  \centering
  \begin{subfigure}[b]{.49\columnwidth}
    \includegraphics[width=\columnwidth]{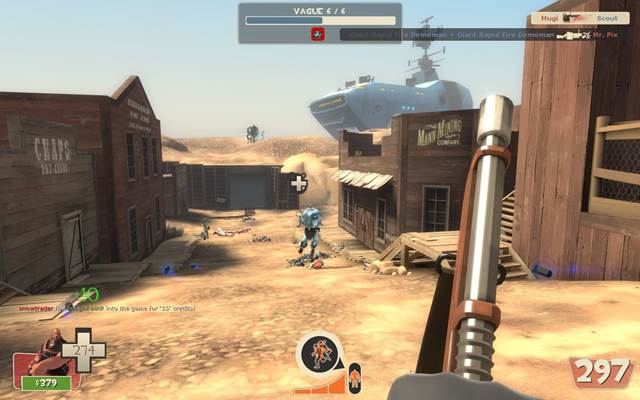}
    \caption{}
  \end{subfigure}
  \begin{subfigure}[b]{.49\columnwidth}
    \includegraphics[width=\columnwidth]{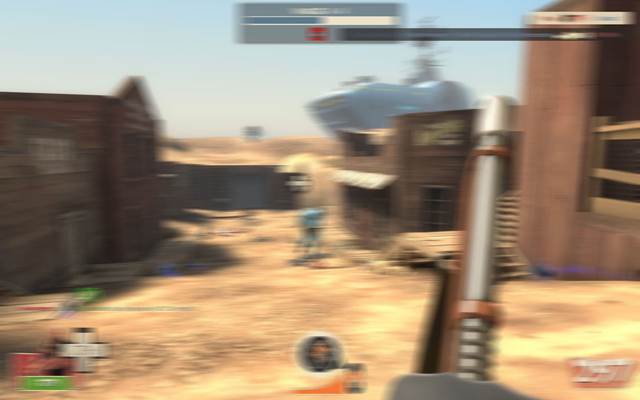}
    \caption{}
  \end{subfigure}
  
  \caption[Example of smear from persistence]{Example of smear from
    persistency. (a)~A rendered scene as seen without head movement. (b)~The
    same scene as perceived by a moving head~(smeared by 2deg).~\footnotemark}

  \label{fig:smearing}
\end{figure}

While motion-to-photon is a well-studied phenomenon in VR, and current
high-quality displays have been improving in this area, another phenomena,
named \emph{smearing}, related to the pixel's persistence has become more
visible.
(Fig.~\ref{fig:smearing} shows an example of how smear is perceived.)
Smearing is caused by an intrinsic interaction between pixel persistence on a
moving display and the Vestibulo-Ocular Reflex~(VOR).
When focusing on one object and rotating our head, the eyes counter-rotate
this movement due to the VOR to keep the image of the object
focused~\cite{regan2017problem}.~\ranote{So what?}

\footnotetext{Image adapted from:
\url{http://blogs.valvesoftware.com/abrash/why-virtual-isnt-real-to-your-brain-judder/}}

Finally, most of the common stereoscopic displays-related distortions are
still present in HMDs. 
For instance, since the current commercially available displays does not
provide eye-tracking technologies accommodation-convergency rivalry is still a
problem for HMDs, and can create several problems such as eyestrain, blurred
image, and misperception of distance, size, depth, or speed of
objects~\cite{patterson2006perceptual}.

\section{Measuring 360-degree content quality}
\label{sec:discussion}


The visual quality of classical images and videos is generally measured by a
global quality index that (ideally) integrates all the possible sources of
distortion into a single value, or a few values.
However, as previously discussed, the sources of distortions in 360-degree
videos are numerous and quite different, and their combination into a single
quality index is far from trivial.

Table~\ref{tbl:artifacts_summary} summarizes the different types of
distortions commonly found in 360-degree video.
In the table, we broadly categorize the artifacts into four categories:
spatial, temporal, stereoscopic, and navigation.
\emph{Spatial artifacts} are those related to still image compression, and can
appear in both images and videos.
\emph{Temporal artifacts} are those related to the temporal evolution of
images and appear only on video.
\emph{Stereoscopic artifacts} are those related to binocular vision.
\emph{Navigation artifacts} are those that only appear while the user
navigates through the scene.

\begin{table*}
\caption{Summary of the visual distortions in 360-degree content.}
\label{tbl:artifacts_summary}
  \scriptsize
  \begin{tabularx}{\textwidth}{c || p{4cm}  p{4cm}  p{4cm} p{3.75cm}}
\toprule
  & \textbf{Capturing} &
    \textbf{Encoding} &
    \textbf{Transmission} &
    \textbf{Display}
  \\
\midrule
    \parbox[t]{2mm}{\vfill \rotatebox[origin=c]{90}{\textbf{Spatial}}}
    &
    Optical distortions (individual cameras):
    \begin{itemize}[leftmargin=*]
      \item blurring by defocus
      \item barrel distortions
      \item pincushion distortions
      \item mustache distortions
      \item noise
      \item chromatic aberrations
    \end{itemize}

    Stitching artifacts:
    \begin{itemize}[leftmargin=*]
      \item discontinuities (e.g., misaligned/broken edges:
      \item missing objects parts
      \item exposure artifacts
      \item black circle / blurred circle
    \end{itemize}

    Blending artifacts:
    \begin{itemize}[leftmargin=*]
      \item Visible color- and luminance- mismatches of regions withing an ODI.
      \item exposure artifacts
      \item visible seams due to color- and luminance- mismatches
      \item ghosting / duplicated objects
    \end{itemize}

    Warping artifacts:
    \begin{itemize}[leftmargin=*]
      \item geometrical distortions / visible deformation of objects
    \end{itemize}
    &
    Projection:
    \begin{itemize}[leftmargin=*]
      \item geometrical distortions
      \item aliasing
        \begin{itemize}
          \item circular pattern aliasing
        \end{itemize}
      \item blurring
      \item ringing
      \item radial pattern close to the poles~(due to oversampling on ERP)
      \item visible seams~(due to sampling)
    \end{itemize}

    Compression:
    \begin{itemize}[leftmargin=*]
      \item blocking
      \begin{itemize}
        \item mosaicing effect
        \item staircase effect
        \item false edges
        \item warped blocking artifacts
        \begin{itemize}
          \item radial blocking pattern~(erp)
          \item perspective projected blocks~(cmp)
          \item identifiable underlying 3D geometry
        \end{itemize}
      \end{itemize}
      \item blurring
      \item ringing
      \item basis pattern effect
      \item color bleeding
      \item visible seams, due to
      \begin{itemize}
        \item color bleeding
        \item crossing faces block transform
        \item loop filter
      \end{itemize}
      \item spatial pattern changes between faces~(cmp)
    \end{itemize}
    &
    Channel distortions:
    \begin{itemize}[leftmargin=*]
      \item data loss
      \item data distortion
    \end{itemize}
    
    Viewport-aware adaptive streaming:
    \begin{itemize}[leftmargin=*]
      \item spatial quality fluctuation
      \item tiling artifacts
    \end{itemize}
    &
    Rendering:
    \begin{itemize}[leftmargin=*]
      \item aliasing
      \item blurring
      \item ringing
    \end{itemize}

    Display limitations:
    \begin{itemize}[leftmargin=*]
      \item Optical distortions
      \begin{itemize}
        \item pincushion distortions
        \item chromatic aberrations
        \item god-rays or flare
        \item \ranote{what else?}
      \end{itemize}
      \item screen-door effect
    \end{itemize}
    \\
\midrule
    \parbox[t]{2mm}{\vfill \rotatebox[origin=c]{90}{\textbf{Temporal}}}
    &
    \begin{itemize}[leftmargin=*]
      \item motion blur
      \item channel mismatch
      \item motion discontinuities
        \begin{itemize}
          \item appearing / disappearing objects
          \item dynamic geometrical distortions
          \item dynamic ghosting
          \item wobbling artifacts
        \end{itemize}
    \end{itemize}
    &
    Compression:
    \begin{itemize}[leftmargin=*]
        \item flickering
        \begin{itemize}
          \item mosquito noise
          \item fine-granularity flickering
          \item coarse-granularity flickering
        \end{itemize}
        \item jerkiness
        \item floating
        \begin{itemize}
          \item texture floating
          \item edge neighborhood floating
        \end{itemize}
      \item flickering seams
      \item spatial pattern changes when crossing the faces
    \end{itemize}
    &
    \begin{itemize}[leftmargin=*]
      \item delay
      \item video freezing
      \item quality fluctuations 
      \item \ranote{Tiling?}
    \end{itemize}
    &
  \\
\hline
  \parbox[t]{2mm}{\vfill \rotatebox[origin=c]{90}{\textbf{Stereoscopy}}}
    &
    \begin{itemize}[leftmargin=*]
      \item depth plane curvature
      \item keystone distortion
      \item cardboard effect
    \end{itemize}
    &
    Projection:
    \begin{itemize}
      \item ghosting~(caused by disocclusion)
    \end{itemize}

    Compression:
    \begin{itemize}[leftmargin=*]
      \item cross-distortions
      \item cardboard effect
    \end{itemize}
    &
    Channel distortions:
    \begin{itemize}[leftmargin=*]
      \item data loss
      \item data distortion~(binocular)
    \end{itemize}

    Viewport-aware adaptive streaming:
    \begin{itemize}[leftmargin=*]
      \item cross distortions~(due to spatial quality fluctuations)
    \end{itemize}
    &
    Display limitations:
    \begin{itemize}[leftmargin=*]
      \item crosstalk as inter-perspective aliasing and ghosting
      \item viewing dependent binocular aliasing
      \item accommodation / convergence rilvary
      \item lattice artifacts
    \end{itemize}

  \\
\hline
    \parbox[t]{2mm}{\vfill \rotatebox[origin=c]{90}{\textbf{Navigation / Head
                                         movement}}}
    &
    &
    &
    Viewport-aware adaptive streaming:
    \begin{itemize}[leftmargin=*]
      \item video freezing
      \item spatial quality fluctuation
      \item quality fluctuation
     \end{itemize}

     Tile-based viewport-aware adaptive streaming:
     \begin{itemize}[leftmargin=*]
       \item spatial quality fluctuations
       \item tile borders
       \item incomplete viewport
     \end{itemize}

    \ranote{what else?}
    &
    Display limitations:
    \begin{itemize}[leftmargin=*]
      \item motion-to-photon delay
      \item smear from persistence
    \end{itemize}
  \\
\bottomrule
\end{tabularx}
\end{table*}



The ultimate way to assess the 360-degree visual quality is through
\emph{subjective} tests, which can shed light on the way the different
distortions interact together.
Such tests, however, are time consuming and expensive.
Thus, \emph{objective} metrics have been proposed for omnidirectional video in
the past few years.
However, it is quite challenging to capture all the effects that impacts the
QoE of 360-degree videos, and much more work remains to be done in this area,
in particular, with regards to perceptually optimized metrics.
The rest of this section presents some of the current approaches for both
objective and subjective quality assessment of 360-degree content and
discusses some of the open research challenges.

\subsection{Objective metrics}

The current global objective quality metrics  for 360-degree content ---such as
standard PSNR and SSIM, viewport-based PSNR/SSIM~\cite{chen2018recent},
S-PSNR~\cite{yu2015framework}, WS-PSNR~\cite{Yule2016WSPSNR}, and
S-SSIM~\cite{chen2018spherical}--- are first attempts to measure the quality
of 360-degree content.
The use of standard image metrics such as PSNR and SSIM directly in the planar
domain is straightforward, but unfortunately:
(1)~they give the same importance to the different parts of the spherical
signal, which besides being sampled very different from classical images, also
have different viewing probabilities~(and then different importance);
(2)~even for traditional images, most of these metrics are kwon for not being
being very good at representing the subjective quality.
The viewport-based PSNR/SSIM metrics apply PSNR and SSIM on the generated
viewports, which is closer to what the users really see.
However, this brings the issue on how to generate representative viewports,
whose number, in theory, can be arbitrary large.

All the above objective metrics, however, fail in \emph{properly considering
the perceptual artifacts} in a 360-degree processing chain, as discussed in
this paper.
For instance, the \emph{visible seams} artifacts due to compression ---which
is usually easy to perceive--- may be hidden in current full-frame objective
metrics, because the samples along the seams are only a small percentage of
the samples in the frame or viewport~\cite{hanhart2018360-degree}.
Thus, objective metrics that detect each artifact reliably and efficiently,
and that can build on the perceptual features of these artifacts are still
necessary.
This paper contribution is a first step towards this direction.

Finally, another important issue today towards the development of perceptually
optimized objective metrics is the \emph{lack of a common quality 360-degree
dataset}~(for both monoscopic and stereoscopic content) to be used for various
dimensions including processing~(fusing, stitching, editing), encoding,
delivery, and rendering/consumption.
It is not clear, for instance, how the few available
datasets~\cite{anne-flore2017measuring,singla2017measuring,xu2017visual,li2018bridge}
cover the visual distortions introduced by state-of-the-art 360-degree
pipelines, and thus how they can be effectively used as benchmark for
perceptual-based quality metrics and for the design of optimized processing
algorithms.
The contribution of this paper can also help in analyzing the current datasets
and on the development of new, more perceptually relevant, ones.

\subsection{Subjective studies}

The lack of common quality 360-degree datasets is also due to the lack of
standardized methodologies for the subjective quality assessment of 360-degree
content, which is still in active debate in the research community.
For instance, through the Immersive Media
Group~(IMG)~\footnote{\url{https://www.its.bldrdoc.gov/vqeg/projects/immersive-media-group.aspx}},
the Video Quality Expert Group~(VQEG) is actively pursuing the development and
standardization of methodologies for the subjective assessment of 360-degree
visual content.
Currently, however, the research community still didn't reach a consensus on
the better methodologies for doing so.

Some recent efforts have been made in adapting subjective methodologies from
classical image/video quality assessment to 360-degree content.
Initial tests have been performed on viewing the rendered viewports on
traditional displays~\cite{boyce2017subjective,vladyslav2016quality}, while
others have been performed using
HMDs~\cite{yu2015framework,upenik2016testbed}.
On the one hand, visualizing the viewports on standard displays lacks the
important immersive features~(increased FoV, magnification of the content,
sense of presence, motion sickness, etc.) that can only be assessed when the
user is wearing an HMD.
On the other hand the adaptation of traditional subjective methods for the
immersive viewing through HMDs is far from trivial because it needs to take
into account at least that:
there are important differences in displays;
the user is immersed in the content;
and that the content can both induce the sense of presence and motion
sickness.

As discussed in Section~\ref{sec:display}, the different displays
specifications, e.g., resolution, supported FoV, etc., may have a direct
impact on the visual quality perceived on subjective studies.
Indeed, as discussed in Section~\ref{sec:display} different HMD lenses may
change how the spatial display resolution is perceived and introduce different
artifacts.
Thus, it is important that during the subjective experiments the specification
of the displays and adaptation of the content for the specific display
Moreover, there is still a lack on cross-device studies that allow researchers
to better understand the impact of the display features on the quality
assessment.

The fact the user is immersed in the content, and free to navigate with 3DoF
the video content completely changes the QoE perspective when compared to
classical subjective studies.
First, by only looking at a fraction of the captured scene at a given time,
the user may not perceive an artifact if he is not looking to the ``right
place''.
Also, since the user is free for turning his head, he will probably not be
able to see the entire scene, and some quality issues may go unnoticed.
Since different people might look at different parts of the content,
\emph{visual attention} and \emph{salient regions} are more important on the
subjective quality assessment of 360-degree content.
Some studies have been considering such importance and datasets have been
proposed to develop these ideas~
Currently, such data have been used mainly for improving streaming content,
but they can~(and should) also be used to improve quality metrics.
Second, the ideal viewing sequence duration, for instance, is not necessarily
the same being standardized for traditional methods, since the user may need
some time to adapt and understand where he is in the content.
\ranote{Long and short time sequences. Add reference to Singla et al. 2017.}

Finally, all the subjective tests performed to the date, including the ones
using HMDs~\cite{yu2015framework,upenik2016testbed}, focus on the overall
signal quality, and do not provide insights on the impact of specific
artifacts and, for instance, how they might cause the user to lose the sense of
presence~(immersion-breaking artifacts).
A better understanding of the impacts of the perceptibility of individual
artifacts and its impacts on user's QoE will only be possible by performing
\emph{psychophysical visual studies}~\cite{wu2013perceptual} specifically
designed for these artifacts, which are still
to appear in the scope of 360-degree content consumed through HMDs. 
We expect that new studies for the other artifacts presented in this paper
will start to appear soon in the literature.

\subsection{Beyond visual quality}

\ranote{What about a discussion on exploratory studies (e.g. focus group) to
better understand the ``unprimed attitudes'', ``feeling'', and ``reactions
towards immersive video?}

Finally, it is important to highlight that visual quality alone is not enough
for measuring QoE in VR.
VR is much broader than just the visual experience, and for a complete
VR~quality framework, besides measuring the visual quality, it is also
necessary to quantify other parameters that have not been discussed in this
study.
For example, VR~Audio, HMD ergonomics~(e.g., weight, weight balance, pressure,
fit and finish, temperature, and overall
hygiene)~\cite{rebelo2019effects,ahram2019user}, user discomfort, and
usability are all important factors in defining a global VR quality of
experience.

Moreover, in this paper we have been mainly concerned visual distortions on
current monoscopic and stereoscopic 360-degree images and videos, which allows
for a 3DoF experience.
New approaches based on multiple 360-degree views, point clouds, and
volumetric videos with potential to support both 3DoF+ and 6DoF are also
expected to appear in the future, and they bring their own issues for visual
quality assessment.
These 

\section{Conclusion}
\label{sec:conclusion}

By reviewing and characterizing the common artifacts in state-of-the-art
end-to-end 360-degree video workflows, this paper contribution is an important
step towards the design of more effective algorithms, applications, and in the
development of perceptual-based quality metrics for 360-degree content~(which
is still an open research problem).
Being aware of the artifacts, understanding their sources, and impact on the
human-visual system can also provide new insights on how to measure, avoid,
and compensate for them.
Indeed, overall, the consideration of the human visual perception in
360-degree video encoder design is an important issue to take into account.

\bibliographystyle{IEEEtran}
\bibliography{IEEEabrv,manuscript}

\begin{thebibliography}{10}
\providecommand{\url}[1]{#1}
\csname url@samestyle\endcsname
\providecommand{\newblock}{\relax}
\providecommand{\bibinfo}[2]{#2}
\providecommand{\BIBentrySTDinterwordspacing}{\spaceskip=0pt\relax}
\providecommand{\BIBentryALTinterwordstretchfactor}{4}
\providecommand{\BIBentryALTinterwordspacing}{\spaceskip=\fontdimen2\font plus
\BIBentryALTinterwordstretchfactor\fontdimen3\font minus
  \fontdimen4\font\relax}
\providecommand{\BIBforeignlanguage}[2]{{%
\expandafter\ifx\csname l@#1\endcsname\relax
\typeout{** WARNING: IEEEtran.bst: No hyphenation pattern has been}%
\typeout{** loaded for the language `#1'. Using the pattern for}%
\typeout{** the default language instead.}%
\else
\language=\csname l@#1\endcsname
\fi
#2}}
\providecommand{\BIBdecl}{\relax}
\BIBdecl

\bibitem{Micusik2004}
B.~Micusik, ``Two view geometry of omnidirectional cameras,'' Ph.D.
  dissertation, Czech Technical University in Prague, 2004.

\bibitem{Chen2018}
\BIBentryALTinterwordspacing
Z.~Chen, Y.~Li, and Y.~Zhang, ``Recent advances in omnidirectional video coding
  for virtual reality: Projection and evaluation,'' \emph{Signal Processing},
  vol. 146, pp. 66 -- 78, 2018. [Online]. Available:
  \url{http://www.sciencedirect.com/science/article/pii/S0165168418300057}
\BIBentrySTDinterwordspacing

\bibitem{DeSimone2016}
F.~D. Simone, P.~Frossard, P.~Wilkins, N.~Birkbeck, and A.~C. Kokaram,
  ``Geometry-driven quantization for omnidirectional image coding,'' in
  \emph{Proc. of the Picture Coding Symposium}, 2016.

\bibitem{petry2015}
B.~Petry and J.~Huber, ``Towards effective interaction with omnidirectional
  videos using immersive virtual reality headsets,'' in \emph{Proceedings of
  the 6th Augmented Human International Conference}.\hskip 1em plus 0.5em minus
  0.4em\relax ACM, 2015, pp. 217--218.

\bibitem{hosseini2017view-aware}
M.~Hosseini, ``View-aware tile-based adaptations in 360 virtual reality video
  streaming,'' in \emph{2017 IEEE Virtual Reality (VR)}, March 2017, pp.
  423--424.

\bibitem{desimone2017omnidirectional}
F.~{De Simone}, P.~Frossard, C.~Brown, N.~Birkbeck, and B.~Adsumilli,
  ``Omnidirectional video communications: new challenges for the quality
  assessment community,'' November 2017.

\bibitem{yuen1998survey}
\BIBentryALTinterwordspacing
M.~Yuen and H.~Wu, ``\BIBforeignlanguage{en}{A survey of hybrid
  {MC}/{DPCM}/{DCT} video coding distortions},''
  \emph{\BIBforeignlanguage{en}{Signal Processing}}, vol.~70, no.~3, pp.
  247--278, Nov. 1998. [Online]. Available:
  \url{http://linkinghub.elsevier.com/retrieve/pii/S0165168498001285}
\BIBentrySTDinterwordspacing

\bibitem{wu2005coding}
M.~Yuen, ``Coding artifacts and visual distortions,'' in \emph{Digital video
  image quality and perceptual coding}, H.~Wu and K.~Rao, Eds.\hskip 1em plus
  0.5em minus 0.4em\relax CRC Press, 2005, pp. 123--158.

\bibitem{unterweger2013compression}
\BIBentryALTinterwordspacing
A.~Unterweger, ``\BIBforeignlanguage{en}{Compression artifacts in modern video
  coding and state-of-the-art means of compensation},'' in
  \emph{\BIBforeignlanguage{en}{Multimedia Networking and Coding}}, R.~A.
  Farrugia and C.~J. Debono, Eds.\hskip 1em plus 0.5em minus 0.4em\relax IGI
  Global, 2013, pp. 28--49. [Online]. Available:
  \url{http://services.igi-global.com/resolvedoi/resolve.aspx?doi=10.4018/978-1-4666-2660-7}
\BIBentrySTDinterwordspacing

\bibitem{zeng2014characterizing}
\BIBentryALTinterwordspacing
K.~Zeng, T.~Zhao, A.~Rehman, and Z.~Wang,
  ``\BIBforeignlanguage{en}{Characterizing perceptual artifacts in compressed
  video streams},'' B.~E. Rogowitz, T.~N. Pappas, and H.~de~Ridder, Eds., Feb.
  2014, p. 90140Q. [Online]. Available:
  \url{http://proceedings.spiedigitallibrary.org/proceeding.aspx?doi=10.1117/12.2043128}
\BIBentrySTDinterwordspacing

\bibitem{meesters_survey_2004}
\BIBentryALTinterwordspacing
L.~Meesters, W.~IJsselsteijn, and P.~Seuntiens, ``\BIBforeignlanguage{en}{A
  {Survey} of {Perceptual} {Evaluations} and {Requirements} of
  {Three}-{Dimensional} {TV}},'' \emph{\BIBforeignlanguage{en}{IEEE
  Transactions on Circuits and Systems for Video Technology}}, vol.~14, no.~3,
  pp. 381--391, Mar. 2004. [Online]. Available:
  \url{http://ieeexplore.ieee.org/document/1273547/}
\BIBentrySTDinterwordspacing

\bibitem{boev2008classification}
\BIBentryALTinterwordspacing
A.~Boev, D.~Hollosi, and A.~Gotchev, ``Classification of stereoscopic
  artefacts,'' \emph{Mobile3DTV Project report, available online at
  http://mobile3dtv. eu/results}, 2008. [Online]. Available:
  \url{http://sp.cs.tut.fi/mobile3dtv/results/tech/D5.1_Mobile3DTV_v1.0.pdf}
\BIBentrySTDinterwordspacing

\bibitem{boev2009stereoscopic}
A.~Boev, A.~Gotchev, and K.~Egiazarian, ``\BIBforeignlanguage{en}{Stereoscopic
  {Artifacts} on {Portable} {Auto}-stereoscopic {Displays}: {What} matters?}''
  p.~6, 2009.

\bibitem{hanhart20133d}
\BIBentryALTinterwordspacing
P.~Hanhart, F.~De~Simone, M.~Rerabek, and T.~Ebrahimi,
  ``\BIBforeignlanguage{en}{3d {Video} {Quality} {Assessment}},'' in
  \emph{\BIBforeignlanguage{en}{Emerging {Technologies} for 3D {Video}}},
  F.~Dufaux, B.~Pesquet-Popescu, and M.~Cagnazzo, Eds.\hskip 1em plus 0.5em
  minus 0.4em\relax Chichester, UK: John Wiley \& Sons, Ltd, Apr. 2013, pp.
  377--391. [Online]. Available:
  \url{http://doi.wiley.com/10.1002/9781118583593.ch19}
\BIBentrySTDinterwordspacing

\bibitem{knorr2017modular}
S.~Knorr, S.~Croci, and A.~Smolic, ``A {Modular} {Scheme} for {Artifact}
  {Detection} in {Stereoscopic} {Omni}-{Directional} {Images},'' in
  \emph{Proceedings of the {Irish} {Machine} {Vision} and {Image} {Processing}
  {Conference}}, 2017.

\bibitem{ebrahimi2017measuring}
T.~Ebrahimi, A.-F. Perrin, C.~Bist, and R.~Cozot, ``Measuring quality of
  omnidirectional high dynamic range content,'' A.~G. Tescher, Ed.\hskip 1em
  plus 0.5em minus 0.4em\relax SPIE, Sep. 2017, p.~38.

\bibitem{schatz2017towards}
\BIBentryALTinterwordspacing
R.~Schatz, A.~Sackl, C.~Timmerer, and B.~Gardlo, ``Towards subjective quality
  of experience assessment for omnidirectional video streaming.''\hskip 1em
  plus 0.5em minus 0.4em\relax IEEE, May 2017, pp. 1--6. [Online]. Available:
  \url{http://ieeexplore.ieee.org/document/7965657/}
\BIBentrySTDinterwordspacing

\bibitem{duan2018perceptual}
H.~Duan, G.~Zhai, X.~Min, Y.~Zhu, Y.~Fang, and X.~Yang,
  ``\BIBforeignlanguage{en}{Perceptual {Quality} {Assessment} of
  {Omnidirectional} {Images}},'' p.~5, 2018.

\bibitem{lim2018vr}
H.-t. {Lim}, H.~G. {Kim}, and Y.~M. {Ro}, ``{VR IQA NET: Deep Virtual Reality
  Image Quality Assessment using Adversarial Learning},'' \emph{ArXiv
  e-prints}, Apr. 2018.

\bibitem{xu2017visual}
M.~Xu, C.~Li, Z.~Wang, and Z.~Chen, ``Visual {Quality} {Assessment} of
  {Panoramic} {Video},'' \emph{arXiv preprint arXiv:1709.06342}, 2017.

\bibitem{zhang2018subjective}
\BIBentryALTinterwordspacing
Y.~Zhang, Y.~Wang, F.~Liu, Z.~Liu, Y.~Li, D.~Yang, and Z.~Chen,
  ``\BIBforeignlanguage{en}{Subjective {Panoramic} {Video} {Quality}
  {Assessment} {Database} for {Coding} {Applications}},''
  \emph{\BIBforeignlanguage{en}{IEEE Transactions on Broadcasting}}, pp. 1--13,
  2018. [Online]. Available:
  \url{https://ieeexplore.ieee.org/document/8350375/}
\BIBentrySTDinterwordspacing

\bibitem{zou2018perceptual}
\BIBentryALTinterwordspacing
W.~Zou, F.~Yang, and S.~Wan, ``\BIBforeignlanguage{en}{Perceptual video quality
  metric for compression artefacts: from two-dimensional to omnidirectional},''
  \emph{\BIBforeignlanguage{en}{IET Image Processing}}, vol.~12, no.~3, pp.
  374--381, Mar. 2018. [Online]. Available:
  \url{http://digital-library.theiet.org/content/journals/10.1049/iet-ipr.2017.0826}
\BIBentrySTDinterwordspacing

\bibitem{yagi1999omnidirectional}
Y.~Yagi, ``Omnidirectional sensing and its applications,'' \emph{IEICE
  Transactions on Information and Systems}, vol.~82, no.~3, pp. 568--579, 1999.

\bibitem{gurrieri2013acquisition}
\BIBentryALTinterwordspacing
L.~E. Gurrieri and E.~Dubois, ``\BIBforeignlanguage{en}{Acquisition of
  omnidirectional stereoscopic images and videos of dynamic scenes: a
  review},'' \emph{\BIBforeignlanguage{en}{Journal of Electronic Imaging}},
  vol.~22, no.~3, p. 030902, Jul. 2013. [Online]. Available:
  \url{http://electronicimaging.spiedigitallibrary.org/article.aspx?doi=10.1117/1.JEI.22.3.030902}
\BIBentrySTDinterwordspacing

\bibitem{szeliski2007image}
R.~Szeliski \emph{et~al.}, ``Image alignment and stitching: A tutorial,''
  \emph{Foundations and Trends{\textregistered} in Computer Graphics and
  Vision}, vol.~2, no.~1, pp. 1--104, 2007.

\bibitem{xu2012panoramic}
W.~Xu, ``\BIBforeignlanguage{en}{Panoramic {Video} {Stitching}},'' Ph.{D}.
  {Thesis}, University of Colorado, Boulder, 2012.

\bibitem{jiang2015video}
\BIBentryALTinterwordspacing
W.~Jiang and J.~Gu, ``\BIBforeignlanguage{en}{Video stitching with
  spatial-temporal content-preserving warping}.''\hskip 1em plus 0.5em minus
  0.4em\relax IEEE, Jun. 2015, pp. 42--48. [Online]. Available:
  \url{http://ieeexplore.ieee.org/document/7301374/}
\BIBentrySTDinterwordspacing

\bibitem{pearson1990map}
F.~Pearson, II, \emph{Map Projections: Theory and Applications}.\hskip 1em plus
  0.5em minus 0.4em\relax CRC press, 1990.

\bibitem{peleg2001omnistereo}
\BIBentryALTinterwordspacing
S.~Peleg, M.~Ben-Ezra, and Y.~Pritch, ``Omnistereo: panoramic stereo imaging,''
  \emph{IEEE Transactions on Pattern Analysis and Machine Intelligence},
  vol.~23, no.~3, pp. 279--290, Mar. 2001. [Online]. Available:
  \url{http://ieeexplore.ieee.org/document/910880/}
\BIBentrySTDinterwordspacing

\bibitem{schroers2018omnistereoscopic}
\BIBentryALTinterwordspacing
C.~Schroers, J.-C. Bazin, and A.~Sorkine-Hornung, ``An omnistereoscopic video
  pipeline for capture and display of real-world vr,'' \emph{ACM Trans.
  Graph.}, vol.~37, no.~3, pp. 37:1--37:13, Aug. 2018. [Online]. Available:
  \url{http://doi.acm.org/10.1145/3225150}
\BIBentrySTDinterwordspacing

\bibitem{anderson2016jump}
\BIBentryALTinterwordspacing
R.~Anderson, D.~Gallup, J.~T. Barron, J.~Kontkanen, N.~Snavely, C.~Hernández,
  S.~Agarwal, and S.~M. Seitz, ``\BIBforeignlanguage{en}{Jump: virtual reality
  video},'' \emph{\BIBforeignlanguage{en}{ACM Transactions on Graphics}},
  vol.~35, no.~6, pp. 1--13, Nov. 2016. [Online]. Available:
  \url{http://dl.acm.org/citation.cfm?doid=2980179.2980257}
\BIBentrySTDinterwordspacing

\bibitem{tan2018360-degree}
\BIBentryALTinterwordspacing
J.~Tan, G.~Cheung, and R.~Ma, ``360-{Degree} {Virtual}-{Reality} {Cameras} for
  the {Masses},'' \emph{IEEE MultiMedia}, vol.~25, no.~1, pp. 87--94, Mar.
  2018. [Online]. Available:
  \url{doi.ieeecomputersociety.org/10.1109/MMUL.2018.011921238}
\BIBentrySTDinterwordspacing

\bibitem{ishiguro1992ods}
H.~Ishiguro, M.~Yamamoto, and S.~Tsuji, ``Omni-directional stereo,'' \emph{IEEE
  Transactions on Pattern Analysis and Machine Intelligence}, vol.~14, no.~2,
  pp. 257--262, Feb 1992.

\bibitem{wu2017dataset}
\BIBentryALTinterwordspacing
C.~Wu, Z.~Tan, Z.~Wang, and S.~Yang, ``A dataset for exploring user behaviors
  in vr spherical video streaming,'' in \emph{Proceedings of the 8th ACM on
  Multimedia Systems Conference}, ser. MMSys'17.\hskip 1em plus 0.5em minus
  0.4em\relax New York, NY, USA: ACM, 2017, pp. 193--198. [Online]. Available:
  \url{http://doi.acm.org/10.1145/3083187.3083210}
\BIBentrySTDinterwordspacing

\bibitem{corbillon2017360-degree}
\BIBentryALTinterwordspacing
X.~Corbillon, F.~De~Simone, and G.~Simon,
  ``\BIBforeignlanguage{en}{360-{Degree} {Video} {Head} {Movement}
  {Dataset}}.''\hskip 1em plus 0.5em minus 0.4em\relax ACM Press, 2017, pp.
  199--204. [Online]. Available:
  \url{http://dl.acm.org/citation.cfm?doid=3083187.3083215}
\BIBentrySTDinterwordspacing

\bibitem{chen2018recent}
\BIBentryALTinterwordspacing
Z.~Chen, Y.~Li, and Y.~Zhang, ``\BIBforeignlanguage{en}{Recent advances in
  omnidirectional video coding for virtual reality: {Projection} and
  evaluation},'' \emph{\BIBforeignlanguage{en}{Signal Processing}}, vol. 146,
  pp. 66--78, May 2018. [Online]. Available:
  \url{http://linkinghub.elsevier.com/retrieve/pii/S0165168418300057}
\BIBentrySTDinterwordspacing

\bibitem{facebook2016cubemap_blog}
\BIBentryALTinterwordspacing
``Next-generation video encoding techniques for 360 video and vr,'' 2016.
  [Online]. Available:
  \url{https://code.fb.com/virtual-reality/next-generation-video-encoding-techniques-for-360-video-and-vr/}
\BIBentrySTDinterwordspacing

\bibitem{GL44spec}
\BIBentryALTinterwordspacing
{Khronos Group}, \emph{{The OpenGL Graphics System: A Specification}}, {Version
  4.4 (Compatibility Profile)}~ed., J.~Leech, Ed., 2014. [Online]. Available:
  \url{https://www.opengl.org/registry/doc/glspec44.compatibility.pdf}
\BIBentrySTDinterwordspacing

\bibitem{dufaux20133d}
\BIBentryALTinterwordspacing
M.~Cagnazzo, B.~Pesquet-Popescu, and F.~Dufaux, ``\BIBforeignlanguage{en}{3d
  {Video} {Representation} and {Formats}},'' in
  \emph{\BIBforeignlanguage{en}{Emerging {Technologies} for 3D {Video}}},
  F.~Dufaux, B.~Pesquet-Popescu, and M.~Cagnazzo, Eds.\hskip 1em plus 0.5em
  minus 0.4em\relax Chichester, UK: John Wiley \& Sons, Ltd, Apr. 2013, pp.
  102--120. [Online]. Available:
  \url{http://doi.wiley.com/10.1002/9781118583593.ch6}
\BIBentrySTDinterwordspacing

\bibitem{stockhammer2011dynamic}
\BIBentryALTinterwordspacing
T.~Stockhammer, ``Dynamic adaptive streaming over http --: Standards and design
  principles,'' in \emph{Proceedings of the Second Annual ACM Conference on
  Multimedia Systems}, ser. MMSys '11.\hskip 1em plus 0.5em minus 0.4em\relax
  New York, NY, USA: ACM, 2011, pp. 133--144. [Online]. Available:
  \url{http://doi.acm.org/10.1145/1943552.1943572}
\BIBentrySTDinterwordspacing

\bibitem{hosseini2016adaptive}
\BIBentryALTinterwordspacing
M.~Hosseini and V.~Swaminathan, ``Adaptive 360 {VR} video streaming: Divide and
  conquer!'' \emph{CoRR}, vol. abs/1609.08729, 2016. [Online]. Available:
  \url{http://arxiv.org/abs/1609.08729}
\BIBentrySTDinterwordspacing

\bibitem{elganainy2016streaming}
\BIBentryALTinterwordspacing
T.~El{-}Ganainy and M.~Hefeeda, ``Streaming virtual reality content,''
  \emph{CoRR}, vol. abs/1612.08350, 2016. [Online]. Available:
  \url{http://arxiv.org/abs/1612.08350}
\BIBentrySTDinterwordspacing

\bibitem{sreedhar2016viewport}
K.~K. Sreedhar, A.~Aminlou, M.~M. Hannuksela, and M.~Gabbouj,
  ``Viewport-adaptive encoding and streaming of 360-degree video for virtual
  reality applications,'' in \emph{2016 IEEE International Symposium on
  Multimedia (ISM)}, Dec 2016, pp. 583--586.

\bibitem{ozcinar2017viewport}
\BIBentryALTinterwordspacing
C.~Ozcinar, A.~D. Abreu, and A.~Smolic, ``Viewport-aware adaptive 360-degree
  video streaming using tiles for virtual reality,'' \emph{CoRR}, vol.
  abs/1711.02386, 2017. [Online]. Available:
  \url{http://arxiv.org/abs/1711.02386}
\BIBentrySTDinterwordspacing

\bibitem{corbillon2017viewport}
X.~Corbillon, G.~Simon, A.~Devlic, and J.~Chakareski, ``Viewport-adaptive
  navigable 360-degree video delivery,'' in \emph{2017 IEEE International
  Conference on Communications (ICC)}, May 2017, pp. 1--7.

\bibitem{graf2017towards}
\BIBentryALTinterwordspacing
M.~Graf, C.~Timmerer, and C.~Mueller, ``Towards bandwidth efficient adaptive
  streaming of omnidirectional video over http: Design, implementation, and
  evaluation,'' in \emph{Proceedings of the 8th ACM on Multimedia Systems
  Conference}, ser. MMSys'17.\hskip 1em plus 0.5em minus 0.4em\relax New York,
  NY, USA: ACM, 2017, pp. 261--271. [Online]. Available:
  \url{http://doi.acm.org/10.1145/3083187.3084016}
\BIBentrySTDinterwordspacing

\bibitem{xie2017360ProbDASH}
\BIBentryALTinterwordspacing
L.~Xie, Z.~Xu, Y.~Ban, X.~Zhang, and Z.~Guo, ``{360ProbDASH}: Improving qoe of
  360 video streaming using tile-based {HTTP} adaptive streaming,'' in
  \emph{Proceedings of the 2017 {ACM} on Multimedia Conference, {MM} 2017,
  Mountain View, CA, USA, October 23-27, 2017}, 2017, pp. 315--323. [Online].
  Available: \url{http://doi.acm.org/10.1145/3123266.3123291}
\BIBentrySTDinterwordspacing

\bibitem{liu2017joint}
K.~Liu, Y.~Liu, J.~Liu, A.~Argyriou, and X.~Yang, ``Joint source encoding and
  networking optimization for panoramic video streaming over lte-a downlink,''
  in \emph{GLOBECOM 2017 - 2017 IEEE Global Communications Conference}, Dec
  2017, pp. 1--7.

\bibitem{ghosh2017rate}
\BIBentryALTinterwordspacing
A.~Ghosh, V.~Aggarwal, and F.~Qian, ``A rate adaptation algorithm for
  tile-based 360-degree video streaming,'' \emph{CoRR}, vol. abs/1704.08215,
  2017. [Online]. Available: \url{http://arxiv.org/abs/1704.08215}
\BIBentrySTDinterwordspacing

\bibitem{zhou2017measurement}
\BIBentryALTinterwordspacing
C.~Zhou, Z.~Li, and Y.~Liu, ``A measurement study of oculus 360 degree video
  streaming,'' in \emph{Proceedings of the 8th ACM on Multimedia Systems
  Conference}, ser. MMSys'17.\hskip 1em plus 0.5em minus 0.4em\relax New York,
  NY, USA: ACM, 2017, pp. 27--37. [Online]. Available:
  \url{http://doi.acm.org/10.1145/3083187.3083190}
\BIBentrySTDinterwordspacing

\bibitem{timmerer2017adaptive}
C.~Timmerer, M.~Graf, and C.~Mueller, ``\BIBforeignlanguage{EN}{Adaptive
  streaming of vr/360-degree immersive media services with high qoe},'' in
  \emph{\BIBforeignlanguage{EN}{2018 NAB Broadcast Engineering and IT
  Conference (BEITC)}}, n.~available, Ed.\hskip 1em plus 0.5em minus
  0.4em\relax Washington DC, USA: National Association of Broadcasters (NAB),
  apr 2017, p.~5.

\bibitem{concolato2017adaptive}
C.~Concolato, J.~L. Feuvre, F.~Denoual, F.~Maze, N.~Ouedraogo, and J.~Taquet,
  ``Adaptive streaming of hevc tiled videos using mpeg-dash,'' \emph{IEEE
  Transactions on Circuits and Systems for Video Technology}, vol.~PP, no.~99,
  pp. 1--1, 2017.

\bibitem{zhou2018effectiveness}
\BIBentryALTinterwordspacing
C.~Zhou, Z.~Li, J.~Osgood, and Y.~Liu, ``On the effectiveness of offset
  projections for 360-degree video streaming,'' \emph{ACM Trans. Multimedia
  Comput. Commun. Appl.}, vol.~14, no.~3s, pp. 62:1--62:24, Jun. 2018.
  [Online]. Available: \url{http://doi.acm.org/10.1145/3209660}
\BIBentrySTDinterwordspacing

\bibitem{ozcinar2017estimation}
C.~Ozcinar, A.~D. Abreu, S.~Knorr, and A.~Smolic, ``Estimation of optimal
  encoding ladders for tiled 360° vr video in adaptive streaming systems,'' in
  \emph{2017 IEEE International Symposium on Multimedia (ISM)}, Dec 2017, pp.
  45--52.

\bibitem{bagnato2012plenoptic}
L.~Bagnato, P.~Frossard, and P.~Vandergheynst, ``Plenoptic spherical
  sampling,'' in \emph{Image Processing (ICIP), 2012 19th IEEE International
  Conference on}.\hskip 1em plus 0.5em minus 0.4em\relax IEEE, 2012, pp.
  357--360.

\bibitem{sauer2018geometry-corrected}
J.~Sauer, M.~Wien, J.~Schneider, and M.~Blaser,
  ``\BIBforeignlanguage{en}{Geometry-{Corrected} {Deblocking} {Filter} for
  360° {Video} {Coding} using {Cube} {Representation}},'' San Francisco, CA,
  2018, p.~5.

\bibitem{desimone2017deformable}
F.~D. Simone, P.~Frossard, N.~Birkbeck, and B.~Adsumilli, ``Deformable
  block-based motion estimation in omnidirectional image sequences,'' in
  \emph{2017 IEEE 19th International Workshop on Multimedia Signal Processing
  (MMSP)}, Oct 2017, pp. 1--6.

\bibitem{naik2018optimized}
\BIBentryALTinterwordspacing
D.~Naik, I.~D.~D. Curcio, and H.~Toukomaa, ``Optimized {Viewport} {Dependent}
  {Streaming} of {Stereoscopic} {Omnidirectional} {Video},'' in
  \emph{Proceedings of the 23rd {Packet} {Video} {Workshop}}, ser. {PV}
  '18.\hskip 1em plus 0.5em minus 0.4em\relax New York, NY, USA: ACM, 2018, pp.
  37--42. [Online]. Available: \url{http://doi.acm.org/10.1145/3210424.3210437}
\BIBentrySTDinterwordspacing

\bibitem{Garcia2014}
M.~N. Garcia, F.~D. Simone, S.~Tavakoli, N.~Staelens, S.~Egger, K.~Brunnstrom,
  and A.~Raake, ``Quality of experience and http adaptive streaming: A review
  of subjective studies,'' in \emph{2014 Sixth International Workshop on
  Quality of Multimedia Experience (QoMEX)}, Sept 2014, pp. 141--146.

\bibitem{Seufert2015}
M.~Seufert, S.~Egger, M.~Slanina, T.~Zinner, T.~Hosfeld, and P.~Tran-Gia, ``A
  survey on quality of experience of http adaptive streaming,'' \emph{IEEE
  Communications Surveys Tutorials}, vol.~17, no.~1, pp. 469--492, Firstquarter
  2015.

\bibitem{hamza2014dash}
A.~Hamza and M.~Hefeeda, ``A dash-based free viewpoint video streaming
  system,'' in \emph{Proceedings of Network and Operating System Support on
  Digital Audio and Video Workshop}.\hskip 1em plus 0.5em minus 0.4em\relax
  ACM, 2014, p.~55.

\bibitem{Schatz2017}
R.~Schatz, A.~Sackl, C.~Timmerer, and B.~Gardlo, ``Towards subjective quality
  of experience assessment for omnidirectional video streaming,'' in \emph{2017
  Ninth International Conference on Quality of Multimedia Experience (QoMEX)},
  May 2017, pp. 1--6.

\bibitem{patterson2006perceptual}
\BIBentryALTinterwordspacing
R.~Patterson, M.~D. Winterbottom, and B.~J. Pierce,
  ``\BIBforeignlanguage{en}{Perceptual {Issues} in the {Use} of
  {Head}-{Mounted} {Visual} {Displays}},'' \emph{\BIBforeignlanguage{en}{Human
  Factors: The Journal of the Human Factors and Ergonomics Society}}, vol.~48,
  no.~3, pp. 555--573, Sep. 2006. [Online]. Available:
  \url{http://journals.sagepub.com/doi/10.1518/001872006778606877}
\BIBentrySTDinterwordspacing

\bibitem{geng2018viewing}
Y.~Geng, J.~Gollier, B.~Wheelwright, F.~Peng, Y.~Sulai, B.~Lewis, N.~Chan,
  W.~S.~T. Lam, A.~Fix, D.~Lanman \emph{et~al.}, ``Viewing optics for immersive
  near-eye displays: pupil swim/size and weight/stray light,'' in \emph{Digital
  Optics for Immersive Displays}, vol. 10676.\hskip 1em plus 0.5em minus
  0.4em\relax International Society for Optics and Photonics, 2018, p. 1067606.

\bibitem{pohl2013improved}
D.~Pohl, G.~S. Johnson, and T.~Bolkart, ``Improved pre-warping for wide angle,
  head mounted displays,'' in \emph{Proceedings of the 19th ACM symposium on
  Virtual reality software and technology}.\hskip 1em plus 0.5em minus
  0.4em\relax ACM, 2013, pp. 259--262.

\bibitem{zhao2017estimating}
\BIBentryALTinterwordspacing
J.~Zhao, R.~S. Allison, M.~Vinnikov, and S.~Jennings,
  ``\BIBforeignlanguage{en}{Estimating the motion-to-photon latency in head
  mounted displays}.''\hskip 1em plus 0.5em minus 0.4em\relax IEEE, 2017, pp.
  313--314. [Online]. Available:
  \url{http://ieeexplore.ieee.org/document/7892302/}
\BIBentrySTDinterwordspacing

\bibitem{regan2017problem}
\BIBentryALTinterwordspacing
M.~Regan and G.~S.~P. Miller, ``\BIBforeignlanguage{en}{The {Problem} of
  {Persistence} with {Rotating} {Displays}},''
  \emph{\BIBforeignlanguage{en}{IEEE Transactions on Visualization and Computer
  Graphics}}, vol.~23, no.~4, pp. 1295--1301, Apr. 2017. [Online]. Available:
  \url{http://ieeexplore.ieee.org/document/7829409/}
\BIBentrySTDinterwordspacing

\bibitem{yu2015framework}
M.~Yu, H.~Lakshman, and B.~Girod, ``A framework to evaluate omnidirectional
  video coding schemes,'' in \emph{Proc. of the IEEE International Symposium on
  Mixed and Augmented Reality}, 2015.

\bibitem{Yule2016WSPSNR}
S.~Yule, A.~Lu, , and Y.~Lu, ``{WS-PSNR} for 360 video objective quality
  evaluation,'' in \emph{MPEG Joint Video Exploration Team}, 2016.

\bibitem{chen2018spherical}
S.~Chen, Y.~Zhang, Y.~Li, Z.~Chen, and Z.~Wang, ``Spherical {Structural}
  {Similarity} {Index} for {Objective} {Omnidirectional} {Video} {Quality}
  {Assessment},'' in \emph{2018 {IEEE} {International} {Conference} on
  {Multimedia} and {Expo} ({ICME})}, Jul. 2018, pp. 1--6.

\bibitem{hanhart2018360-degree}
P.~Hanhart, Y.~He, Y.~Ye, J.~Boyce, Z.~Deng, and L.~Xu,
  ``\BIBforeignlanguage{en}{360-{Degree} {Video} {Quality} {Evaluation}},''
  p.~5, 2018.

\bibitem{anne-flore2017measuring}
\BIBentryALTinterwordspacing
R.~C. T.~E. Anne-Flore~Perrin, Cambodge~Bist, ``Measuring quality of
  omnidirectional high dynamic range content,'' vol. 10396, 2017, pp. 10\,396
  -- 10\,396 -- 18. [Online]. Available:
  \url{https://doi.org/10.1117/12.2275146}
\BIBentrySTDinterwordspacing

\bibitem{singla2017measuring}
\BIBentryALTinterwordspacing
A.~Singla, S.~Fremerey, W.~Robitza, and A.~Raake,
  ``\BIBforeignlanguage{en}{Measuring and comparing {QoE} and simulator
  sickness of omnidirectional videos in different head mounted displays},'' in
  \emph{\BIBforeignlanguage{en}{2017 {Ninth} {International} {Conference} on
  {Quality} of {Multimedia} {Experience} ({QoMEX})}}.\hskip 1em plus 0.5em
  minus 0.4em\relax Erfurt, Germany: IEEE, May 2017, pp. 1--6. [Online].
  Available: \url{http://ieeexplore.ieee.org/document/7965658/}
\BIBentrySTDinterwordspacing

\bibitem{li2018bridge}
\BIBentryALTinterwordspacing
C.~Li, M.~Xu, X.~Du, and Z.~Wang, ``Bridge the gap between vqa and human
  behavior on omnidirectional video: A large-scale dataset and a deep learning
  model,'' in \emph{Proceedings of the 26th ACM International Conference on
  Multimedia}, ser. MM '18.\hskip 1em plus 0.5em minus 0.4em\relax New York,
  NY, USA: ACM, 2018, pp. 932--940. [Online]. Available:
  \url{http://doi.acm.org/10.1145/3240508.3240581}
\BIBentrySTDinterwordspacing

\bibitem{boyce2017subjective}
J.~Boyce, E.~Alshina, and Z.~Deng, ``Subjective testing method for 360°
  {Video} projection formats using {HEVC},'' ISO/IEC, Tech. Rep. N16892, Apr.
  2017.

\bibitem{vladyslav2016quality}
\BIBentryALTinterwordspacing
J.~H.~P. Vladyslav~Zakharchenko, Kwang Pyo~Choi, ``Quality metric for spherical
  panoramic video,'' vol. 9970, 2016, pp. 9970--9970--9. [Online]. Available:
  \url{https://doi.org/10.1117/12.2235885}
\BIBentrySTDinterwordspacing

\bibitem{upenik2016testbed}
E.~Upenik, M.~Řeřábek, and T.~Ebrahimi, ``Testbed for subjective evaluation
  of omnidirectional visual content,'' in \emph{2016 {Picture} {Coding}
  {Symposium} ({PCS})}, Dec. 2016, pp. 1--5.

\bibitem{wu2013perceptual}
\BIBentryALTinterwordspacing
H.~R. Wu, A.~R. Reibman, W.~Lin, F.~Pereira, and S.~S. Hemami,
  ``\BIBforeignlanguage{en}{Perceptual {Visual} {Signal} {Compression} and
  {Transmission}},'' \emph{\BIBforeignlanguage{en}{Proceedings of the IEEE}},
  vol. 101, no.~9, pp. 2025--2043, Sep. 2013. [Online]. Available:
  \url{http://ieeexplore.ieee.org/document/6575139/}
\BIBentrySTDinterwordspacing

\bibitem{rebelo2019effects}
\BIBentryALTinterwordspacing
Y.~Yan, K.~Chen, Y.~Xie, Y.~Song, and Y.~Liu, ``The {Effects} of {Weight} on
  {Comfort} of {Virtual} {Reality} {Devices},'' in \emph{Advances in
  {Ergonomics} in {Design}}, F.~Rebelo and M.~M. Soares, Eds.\hskip 1em plus
  0.5em minus 0.4em\relax Cham: Springer International Publishing, 2019, vol.
  777, pp. 239--248. [Online]. Available:
  \url{http://link.springer.com/10.1007/978-3-319-94706-8_27}
\BIBentrySTDinterwordspacing

\bibitem{ahram2019user}
\BIBentryALTinterwordspacing
J.~Zhuang, Y.~Liu, Y.~Jia, and Y.~Huang, ``User {Discomfort} {Evaluation}
  {Research} on the {Weight} and {Wearing} {Mode} of {Head}-{Wearable}
  {Device},'' in \emph{Advances in {Human} {Factors} in {Wearable}
  {Technologies} and {Game} {Design}}, T.~Z. Ahram, Ed.\hskip 1em plus 0.5em
  minus 0.4em\relax Cham: Springer International Publishing, 2019, vol. 795,
  pp. 98--110. [Online]. Available:
  \url{http://link.springer.com/10.1007/978-3-319-94619-1_10}
\BIBentrySTDinterwordspacing

\end{thebibliography}

\end{document}